\documentclass[onecolumn,citeautoscript,superscriptaddress,notitlepage]{revtex4-1}     
\usepackage{amsmath,amssymb,mathrsfs,bm,setspace,xspace,soul,empheq}  
\usepackage{graphicx}
\usepackage{physics} 
\usepackage{float}  
\usepackage{comment} 
\usepackage{siunitx}    
\usepackage[tight]{subfigure}       
\usepackage{braket}
\usepackage{color}  
\usepackage{dcolumn}
\usepackage{multirow}            
\usepackage{geometry}         
\usepackage{tabularx}
\usepackage[colorlinks=true]{hyperref}   
\hypersetup{ 
    bookmarks=true,         
    unicode=false,          
    pdftoolbar=true,        
    pdfmenubar=true,        
    pdffitwindow=false,     
    pdfstartview={FitH},    
    pdftitle={My title},    
    pdfauthor={Author},     
    pdfsubject={Subject},   
    pdfcreator={Creator},   
    pdfproducer={Producer}, 
    pdfkeywords={keyword1} {key2} {key3}, 
    pdfnewwindow=true,      
    colorlinks=true,       
    linkcolor=magenta, 
    citecolor=blue,        
    filecolor=magenta,      
    urlcolor=cyan           
} 

\renewcommand{\abs}[1]{\left| #1 \right|}

\newcommand{\beq}{\begin{equation}}
\newcommand{\eeq}{\end{equation}}
\newcommand{\ba}{\begin{array}{ccc}}
\newcommand{\ea}{\end{array}}

\def\bea{\begin{eqnarray}}
\def\eea{\end{eqnarray}}
\newcommand{\bml}{\begin{multline}}

\newcommand{\eeqm}{\end{multline}}
\newcommand{\bsp}{\begin{split}}
\newcommand{\esp}{\end{split}}

\renewcommand{\b}[1]{{\bf #1}}

\newcommand{\mc}{\mathcal}

\newcommand{\ts}{\thinspace{}}

\newcommand{\approptoinn}[2]{\mathrel{\vcenter{
  \offinterlineskip\halign{\hfil$##$\cr
    #1\propto\cr\noalign{\kern2pt}#1\sim\cr\noalign{\kern-2pt}}}}}

\begin{document}  

\title{Geometric entanglement in integer quantum Hall states}   

\author{Benoit Sirois}
\address{Theoretical Physics, Oxford University, 1 Keble Road, Oxford OX1 3NP, UK}
\address{D\'epartement de Physique, Universit\'e de Montr\'eal, Montr\'eal, Qu\'ebec, H3C 3J7, Canada}
\author{Lucie Maude Fournier}
\author{Julien Leduc}
\address{D\'epartement de Physique, Universit\'e de Montr\'eal, Montr\'eal, Qu\'ebec, H3C 3J7, Canada}
\author{William Witczak-Krempa}
\address{D\'epartement de Physique, Universit\'e de Montr\'eal, Montr\'eal, Qu\'ebec, H3C 3J7, Canada}
\address{Centre de Recherches Math\'ematiques, Universit\'e de Montr\'eal; P.O. Box 6128, Centre-ville Station; Montr\'eal (Qu\'ebec), H3C 3J7, Canada}
\address{Regroupement Qu\'eb\'ecois sur les Mat\'eriaux de Pointe (RQMP)}

\begin{abstract}  
\begin{center}
\textbf{Abstract}
\end{center}
We study the quantum entanglement structure of integer quantum Hall states via the reduced density matrix of spatial subregions. In particular, we examine the eigenstates, spectrum and entanglement entropy (EE) of the density matrix for various ground and excited states, with or without mass anisotropy.
We focus on an important class of regions that contain sharp corners or cusps, leading to a geometric angle-dependent contribution to the EE. We unravel surprising relations by comparing this corner term at different fillings. We further find that the corner term, when properly normalized, has nearly the same angle dependence as numerous conformal field theories (CFTs) in two spatial dimensions, which hints at a broader structure.
In fact, the Hall corner term is found to obey bounds that were previously obtained for CFTs. 
In addition, the low-lying entanglement spectrum and the corresponding eigenfunctions reveal ``excitations'' localized near corners.
Finally, we present an outlook for fractional quantum Hall states.     
\end{abstract}

\date{\today}   
\maketitle      
\tableofcontents       

\section{Introduction}    
Restricting observations to a spatial subregion of a quantum system, such as as a cold two-dimensional electron gas (2DEG), gives information about the entire system due to the presence of entanglement. 
Rather than studying specific observables localized in the region, one can examine the reduced density matrix, which is obtained by taking the full density matrix and tracing out degrees of freedom outside the region of interest.
Due to the large amount of information stored in that reduced density matrix, it is often advantageous to to study parts of it, such as subset of its eigenstates, spectrum (called the entanglement spectrum), and more simply, the entanglement entropy (EE). The latter is a positive number that, heuristically speaking, quantifies how much entanglement exists between a region and its complement.\footnote{This is true at sufficiently low temperatures.}
 
The entanglement spectrum and EE have been particularly useful in revealing the topological properties of two-dimensional (2D) quantum systems~\cite{Kitaev2006,Levin2006,Dong2008,Li2008,FradkinBook}, such as quantum Hall states. Indeed, the topological EE, which depends on the topology of the subregion not its geometry, gives insight about the anyons present in a topologically ordered state. The shape or geometrical dependence of the reduced density matrix also contains rich information about the system. The geometrical aspects have been particularly studied for the ground states of gapless quantum systems, such as conformal field theories,
through the EE~\cite{Fradkin2006,solodukhin_entanglement_2008,Casini_rev_2009,Nishioka_2009,Myers2010,FradkinBook,Kallin2013,Kallin2014,Bueno2015,Faulkner2016,laflorencie2016,BosonsFermions,Chen2017,Bueno2016,WWK2019}.                   
In this work, we analyze the geometrical properties of the reduced density matrix for a particularly simple class of topological states, the integer quantum Hall (IQH) states. Although idealized IQH wavefunctions represent non-interacting fermions, they nevertheless possess a rich spatial entanglement structure. As we shall show, certain entanglement properties of IQH states   
closely resemble those of strongly interacting quantum critical systems in 2 spatial dimensions.          

\subsection{Entanglement entropy in quantum Hall states}
More precisely, the von Neumann EE of a subregion $A$ in a state described by a density matrix $\rho$ is given by 
  $S(A)=-\Tr_{A}\rho_A \ln \rho_A$, where $\rho_A$ is the reduced density matrix of $A$ obtained by tracing out degrees of freedom in the complement $A^c$: $\rho_{A}=\Tr_{A^c} \rho$. In this work, we consider pure states,
  $\rho=|\psi\rangle\langle \psi|$.
The EE for a spatial bipartition of a quantum Hall state described by a trial wavefunction at filling $\nu$, an example being the
electronic Laughlin state at $\nu=1/3$, should take the general form 
\begin{align} \label{eq:arealaw} 
  S(A) = c\frac{L_A}{\ell_B} - \gamma_{\rm top} - \gamma_{\rm geo}  + \dotsb
\end{align}
where $L_A$ is the perimeter of subregion $A$, which we take to be much larger than the magnetic length, $L_A\gg \ell_B$, and the ellipsis denotes terms
that vanish at large $L_A/\ell_B$.
The first term is the area (or boundary) law that is generally present for the low energy states of local Hamiltonians without a finite Fermi surface.
Here, the magnetic length $\ell_B$ plays the role of the microscopic (UV) length scale. In contrast, in a lattice model this role would be played by the lattice spacing,
whereas a UV cutoff would appear in a continuum quantum field theory. 
Let us now examine the subleading corrections parametrized by $\gamma$, as these contain more useful information about the state. 
First, $\gamma_{\rm top}\geq 0$ is the universal topological contribution arising from the topological order associated with the gapped phase~\cite{Kitaev2006,Levin2006,Dong2008,Li2008,FradkinBook}.
It detects the presence of anyon excitations, and is thus absent at integer filling for fermions.   
As its name suggests, $\gamma_{\rm top}$ does not dependent on the geometry of subregion $A$. This term has been widely studied, and can
be obtained using topological quantum field theory (TQFT). For example, for Laughlin states at frational filling $\nu=1/m$, with $m$ odd, it is
$\gamma_{\rm top}=\tfrac12 \ln m$~\cite{Kitaev2006}.
The next term, $\gamma_{\rm geo}$, is a geometric contribution. For Laughlin states, it is universal in the sense that, just as $\gamma_{\rm top}$,  
it does not depend on $\ell_B$ or any other scale. It is a pure number that only depends on the shape of subregion $A$ and the state under consideration. 

An important class of shapes has corners or cusps, such as a triangle or a square. 
The EE of such non-smooth subregions has been extensively studied in the groundstates of gapless Hamiltonians, such as 
conformal field theories~\cite{Casini_rev_2009,Hirata07,Kallin2013,Kallin2014,Miles2014,Bueno2015,BuenoMyers2015,Faulkner2016,Bueno2016,Whitsitt2017,WWK2019,BCW2019}. It was found that the subleading correction $\gamma$ contains 
a contribution diverging logarithmically   
with the perimeter, $\sum_i a(\theta_i)\ln(L_A/\epsilon)$, where $\epsilon$ is  
a short-distance cutoff. The prefactor $\sum_i a(\theta_i)$ depends on the geometry of $A$ through the angles of its corners, $\theta_i$.     
In a variety of states, it was shown that the \emph{corner function}
$a(\theta)$ captures key information about long-distance physics of the quantum critical state, and shows surprising universality.
As a concrete example, when the corner is nearly smooth $\theta\approx \pi$, the corner function yields the stress tensor central charge $C_T$
of conformal field theories in two spatial dimensions~\cite{Bueno2015,BuenoMyers2015,Faulkner2016}. This central charge controls the two-point function of the stress tensor (thus of the energy density), as well
as the finite-frequency shear viscosity.   

In groundstates of gapped systems, like quantum Hall states, the logarithmic divergence is cut off by the finite correlation length.
In that case, we expect that a polygon shaped region will yield the following subleading term:
\begin{align}  \label{eq:cornercontribution}
  \gamma_{\rm geo}= \sum_i a(\theta_i)
\end{align}
that does not diverge with the size of region $A$, in contrast to critical states. 
For trial wavefunctions such as the Laughlin states, the corresponding corner function $a(\theta)$ is a pure number independent of all scales.
For the quantum critical states described above, this pure number is multiplied by a logarithm $a(\theta)\ln(L_A/\epsilon)$, which preserves the cutoff ($\epsilon$) independence of $a(\theta)$.
The function $a(\theta)$ has been computed numerically for the integer quantum Hall state at $\nu=1$~\cite{SierraPolygonal}.  
However, the physical information encoded by this function remains unknown, even in the simplest case at $\nu=1$. It has yet to be computed at 
fractional fillings.
\begin{figure}[t] 
\centering
\includegraphics[scale=0.8]{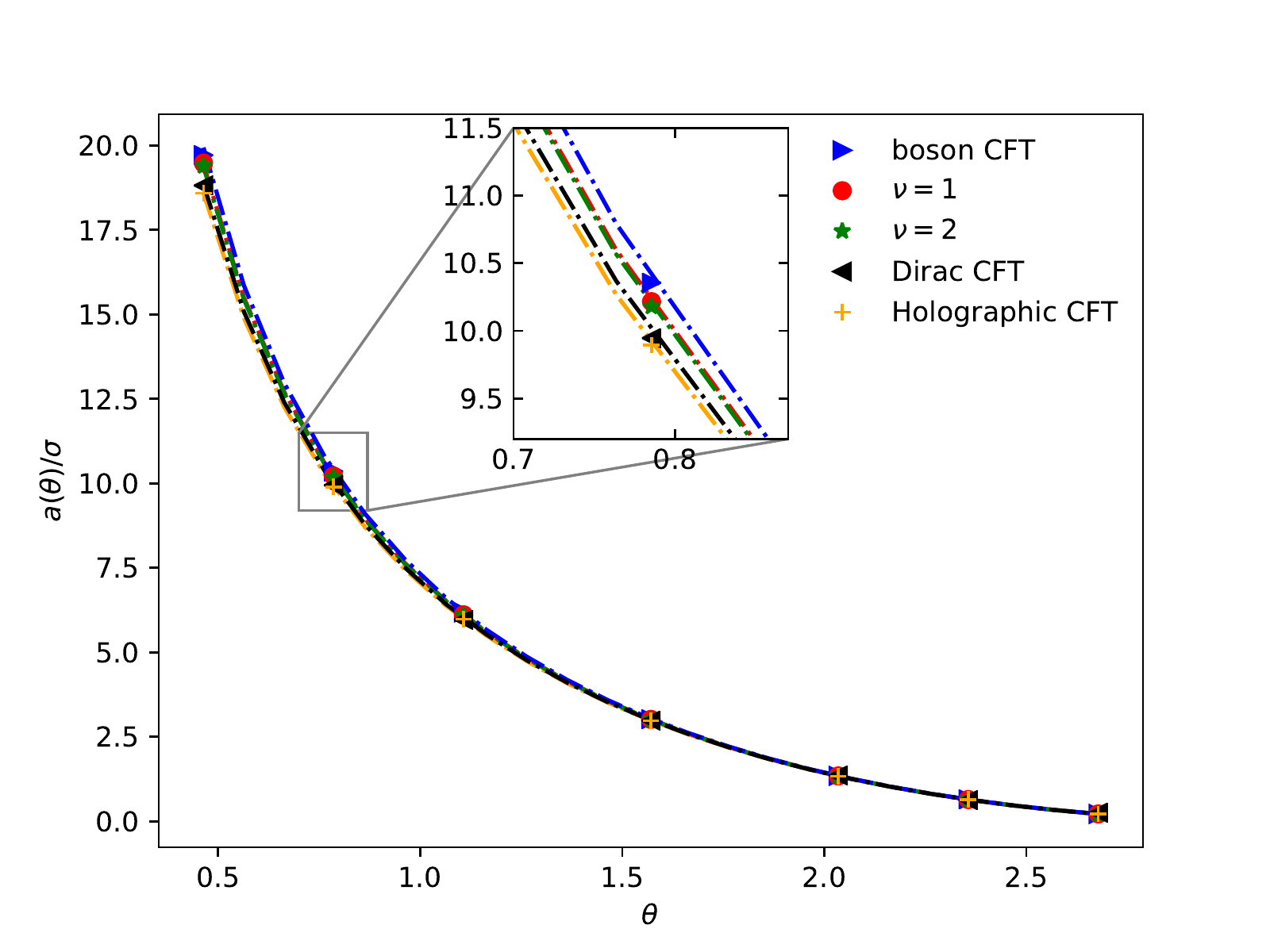}
\caption{\label{figtan3}Comparison of corner functions for various systems in two spatial dimensions, normalized by the smooth-limit coefficient $\sigma$. The integer quantum Hall ground states at fillings $\nu=1,2$ are computed in this work.  
  The boson conformal field  theory (CFT) corresponds to non-interacting massless relativistic bosons~\cite{BosonsFermions}, which have the same corner function as the large-$N$ $O(N)$ Wilson-Fisher fixed point~\cite{Whitsitt2017}.
  The Dirac CFT is a theory of massless Dirac fermions~\cite{BosonsFermions}, while the holographic CFT corresponds to a strongly interacting supersymmetric CFT described by the AdS/CFT correspondence.
  Data markers correspond to lattice simulations (except for the holographic CFT~\cite{Hirata07}), while continuous lines correspond to the ansatz Eq.\ts(\ref{Twist}), except for the boson and Dirac CFT where a more precise
  ansatz is used~\cite{BosonsFermions}.} 
\end{figure}       

In this work, we revisit the calculation of $a(\theta)$ at $\nu=1$ using analytical insights as well as high-precision numerics.   
We also study the entire reduced density matrix $\rho_A$ through its Schmidt spectrum and eigenstates. 
We further extend our calculations to the groundstate at filling $\nu=2$, as well as to the excited state at $\nu=1$ obtained by filling the
first Landau Level and leaving the LLL unoccupied. When only the 1st LL is occupied (0th level empty), 
we find surprising relations to the results for the groundstate at $\nu=1$, helping shed light on the 
physics encoded in the corner function. We push the comparison further by comparing the corner function to the one obtained for various conformal field theories in two spatial dimensions, including gapless Dirac fermions. We find that the Hall corner function, when properly normalized, has a surprisingly close shape dependence to the conformal theories, as shown in Fig.~\ref{figtan3}. 
We also examine the role of anisotropy on the EE, and show that it strongly affects its shape dependence.
Going beyond the EE, we analyze the low-lying entanglement spectrum and the corresponding eigenfunctions, which reveal ``excitations'' localized near corners.
Finally, we study a different type of corner where two tips touch at a point (``hourglass''), and we extract a universal quantity via the mutual information.

The rest of the paper is organized as follows: Section~\ref{sec:ent-IQH} describes how to obtain the reduced density matrix, entanglement spectrum and EE for IQH states. Section~\ref{sec:corner} contains the results for these quantities for simple regions that contain a corner. Section~\ref{secC} describes the effects of mass anisotropy on the EE. Section~\ref{sec:tip-touch} studies a new type of region where two corners touch at a point. This geometry can be used to to define a quantum mutual information that is independent on the microscopic information of the IQH states. Finally, in Section~\ref{sec:concl} we summarize our main findings and present an outlook for interacting systems, including fractional quantum Hall states. Appendices~\ref{AppendixA} and \ref{A:prec} provide detailed information regarding our numerical results.  

\section{Entanglement in integer quantum Hall states} \label{sec:ent-IQH} 
The IQH system can be described by considering the following single-electron Hamiltonian in the Landau gauge:
\begin{align}
H=\frac{p_{x}^{2}}{2m_e}+\frac{\left(p_{y}+eBx\right)^{2}}{2m_e}
\label{h}
\end{align}
where $m_e$ is the effective mass of the electrons. On a cylinder of circumference $L_y$ (shown in Fig.~\ref{schema}), the 
eigenstates of $H$ are the usual Landau level (LL) wavefunctions of energy $E_n=\hbar \omega_c \left(n+\frac{1}{2}\right)$: 
\begin{align}
\phi_{n,k}(x,y) = d_n\,e^{ik y} H_{n}&(x+k \ell_B^2) \exp\left(-\frac{\left(x+k \ell_B^2\right)^{2}}{2\ell_B^2}\right)\,,\quad n=0,1,2,\dots
\label{LL}
\end{align} 
The $y$-periodicity leads to discrete wavevectors $k=2 \pi m/ L_y$ with $m \in \mathbb{Z}$. 
The key scale of the problem is the magnetic length $\ell_B=\sqrt{\hbar /eB}$, and the cyclotron frequency $\omega_c=e B/m_e$ gives
the gap betwee LLs. $H_n$ are Hermite polynomials, and the normalization coefficient is $d_n=\pi^{-1/4}/\sqrt{2^n n!\ell_B L_y}.$
For IQH states, the system's wavefunction is obtained by entirely filling one or more LLs with electrons of every $y$-momentum $k$.     
We will consider the groundstates at $\nu=1$ and $2$. We will also 
study a special excited eigenstate at $\nu=1$ where every electron is in the $n=1$ LL, which we will call the 1st LL excited state.
In our analysis, we shall set $\ell_B=1$. 

We want to study the reduced density matrix of these states for various subregions $A$ of the cylinder. In particular, we will examine the entanglement spectrum, EE, and
eigenstates of the reduced density matrix $\rho_A$.
Since we deal with non-interacting electrons, we can use the method developed in \cite{Peschel}. The eigenvalues of $\rho_A$ can be computed from the eigenvalues of the correlation function $C_{\mathbf{r},\mathbf{r'}}=\braket{c^{\dagger}_{\mathbf{r}}c_{\mathbf{r'}}}$ restricted to subregion $A$, where averages are computed in the state of the total system $A \cup A^c$. For our IQH states, it is possible to discretize this eigenvalue problem, as shown in \cite{SierraPolygonal}, by diagonalizing a block matrix given by  
\begin{align} \label{eq:F-mat}
  \mathcal{F}^{(n,n')}_{k,k'}(A)=\int_A d^2\mathbf{r}\, \phi_{n,k}(\mathbf{r}) \phi_{n',k'}^*(\mathbf{r})\,,
\end{align}
where for each pair of momenta $\{ k ,k'\}$ one has a block corresponding to the occupied LLs. In the simplest case of the IQH at $\nu=1$, the block is only a number. For the IQH at $\nu=2$ we
have a 2-by-2 block corresponding to $n=0,1$; the $(0,1)$ and $(1,0)$ off-diagonal elements measure the overlap in $A$ between electrons in the LLL and the 1st LL.  
For numerical computations to be possible, one needs to truncate the infinite-dimensional $\mathcal{F}$ matrix. The natural way to do so is to work with matrices with $\{k, k'\}$ smaller than a certain cutoff. This amounts
to considering only states $\phi_{n,k}$ centered not too far from the cut. Increasing the cutoff should then lead to convergent results since far away electrons contribute negligibly to the entanglement between $A$ and $A^c$.  

From the discussion above, we understand that the spectrum of the correlator $C$ is of great importance. It is actually directly related to the entanglement spectrum (ES) which is defined as the spectrum of $-\ln \rho_A$ (up to a shift of the zero of the spectrum). Indeed, we can first relate the ES to the \emph{single-particle} spectrum of the entanglement Hamiltonian $H_A$ defined as $\rho_A = \frac{1}{Z} \exp(-H_A)$ \cite{ES}, which in the case of free fermions is a free-fermion quadratic Hamiltonian restricted to subregion $A$:
$H_A = \sum\limits_{\mathbf{r},\mathbf{r'} \in A} h_{\mathbf{r}, \mathbf{r'}} c^{\dagger}_{\mathbf{r}} c_{\mathbf{r'}}$. The eigenvalues $\epsilon$ of $h$ (which we refer to as the spectrum of $H_A$) can be obtained from the eigenvalues of the correlation matrix by the relation $h^T = \ln\left(\frac{1-C}{C}\right)$ \cite{Peschel}. In this work, we numerically compute the eigenvalues $\lambda$ of the correlation matrix $C$. The relation between the two spectra is then given by the matrix relation given just above, or equivalently by: $\lambda=\frac{1}{1+\text{e}^{\epsilon}}=n_F\left(\epsilon\right)$, where $n_F$ is the Fermi-Dirac distribution.
We will from now on refer to the single-particle spectrum of $H_A$ as the ES, since one can reconstruct the full spectrum of $\rho_A$ from the eigenvalues of $h$.

\section{Corner entanglement}\label{sec:corner}

\begin{figure}[t] 
\centering
\includegraphics[scale=0.70]{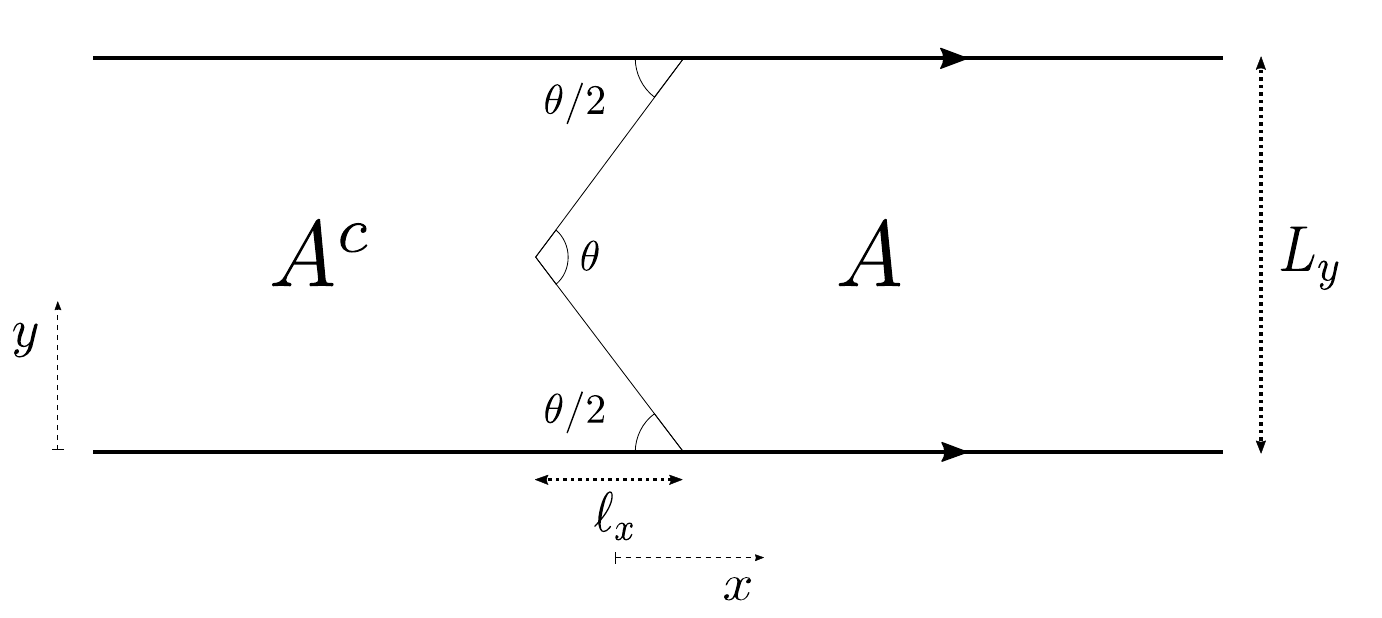}
\caption{\label{schema}The ``arrow-shaped'' subregion $A$ and its complement in an infinite cylinder of circumference $L_y$. $A$ has two corners of angles $\theta$ and $2\pi-\theta$.
  The arrows on the top and bottom of the drawn rectangle indicate the equivalent points. } 
\end{figure}

We are interested in computing the ES and various EEs for IQH states (in which $\gamma_{\rm top}$ vanishes), for domains $A$ with non-smooth boundaries,
and with a perimeter that far exceeds the magnetic length, $L_A/\ell_B\gg 1$. 
  In contrast, for smooth regions (where smoothness is defined relative to the scale $\ell_B$) the residual part of the EE $\gamma_{\rm geo}$ vanishes~\cite{Rodriguez2009,Estienne2019}, as can bee seen for a flat cut on the
  cylinder (see below). This makes the corner geometry even more important for the states under consideration.    
We will calculate the corner contributions to the EE and ES of the arrow-shaped subregion of an infinite cylinder of circumference $L_y$ presented in Fig.\ts\ref{schema}.
Note that due to the periodicity in the $y$ direction, $A$ contains two corners of angles $\theta$ and $2\pi-\theta$.  

\subsection{Entanglement spectrum} 
\begin{figure}
\centering
\subfigure[\label{spectre1} Entanglement spectra in the $\nu = 1$ state.]{\includegraphics[width=.48\textwidth]{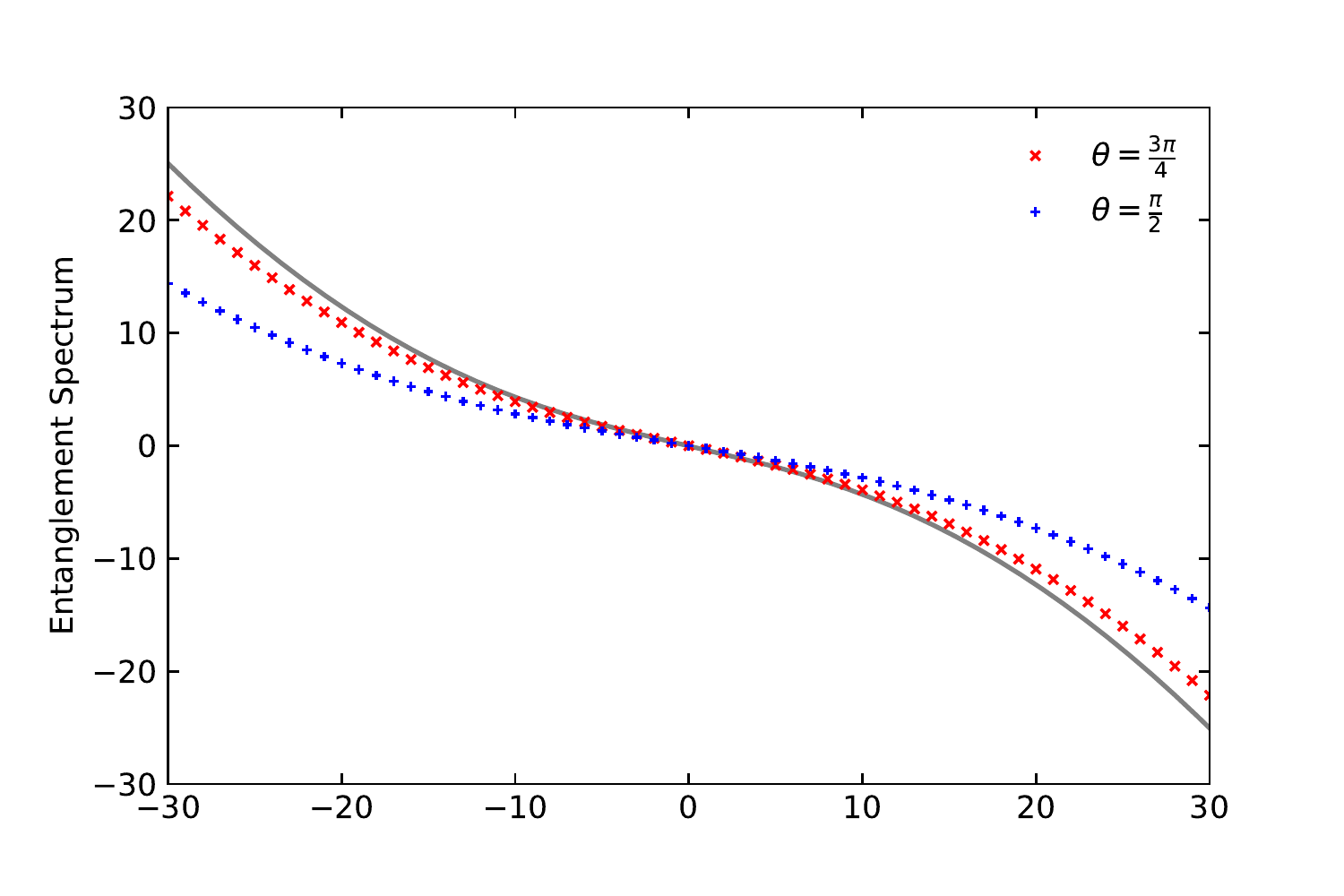}}
\subfigure[\label{spectre2} Entanglement spectra in the filled $1^{st}$ LL excited state.]{\includegraphics[width=.48\textwidth]{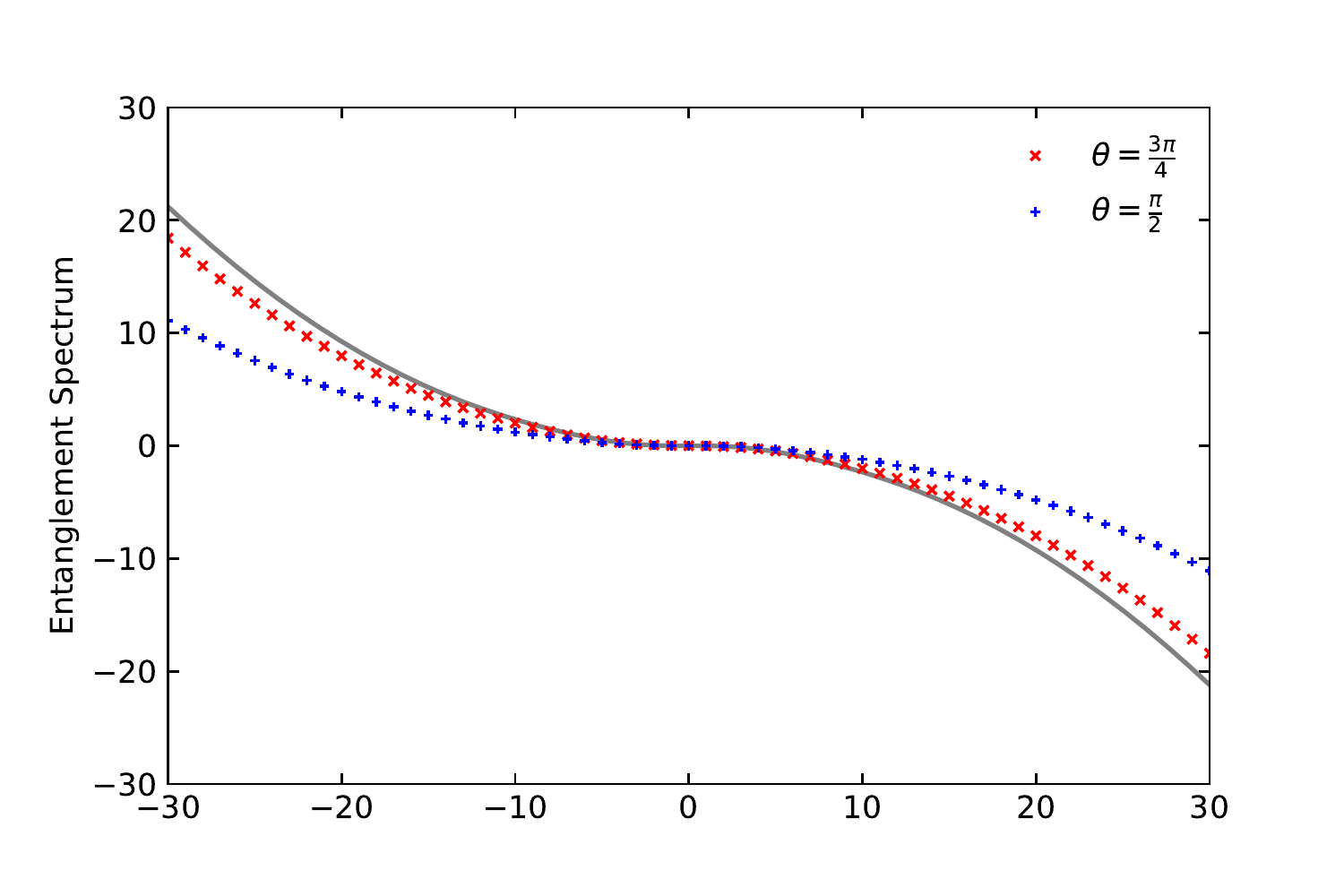}}  
\caption{\label{spectres}Entanglement spectra. The continuous lines are the spectra associated to a smooth cut, while the discrete points correspond to arrow-shaped domains of corner angle $\theta$.
  For these panels and in Fig.~\ref{spectre12} and \ref{spectre12ns}, the sign convention of \cite{SierraPolygonal} is used for the wavefunctions.}
\end{figure}

As can be seen in Fig.\ts\ref{spectres}, the presence of corners results in a deformation of the ES compared to that of a smooth cut for the $\nu = 1$ and filled $1^{st}$ LL states. 
In the case of a smooth cut, an analytical formula exists for the ES of the $\nu = 1$ state \cite{SierraLisse}, and, in general, for all ES of filled $n^{th}$ LL states. For a smooth cut in the $y$ direction, the spectrum can be expressed as a function of the $y$-momentum $k$. At small $k$, we can easily show that an even $n$ (including $n=0$) results in a linear dispersion $\epsilon(k) \propto k$, whereas, for an odd $n$, we have $\epsilon(k) \propto k^{3}$.
For the simplest case of the $\nu=1$ IQH ground state ($n=0$), the linear dispersion matches the dispersion of a chiral mode on a physical edge, in agreement with the connection between entanglement cuts and physical edges~\cite{Li2008}. 
For the 1st LL excited state (all fermions have $n=1$ wavefunctions), the dispersion is cubic at small $k$, and thus does not have the expected dispersion for the corresponding physical edge mode.    

\begin{figure}
\centering
\subfigure[\label{spectre12} Smooth subregion.]{\includegraphics[width = 0.45\textwidth]{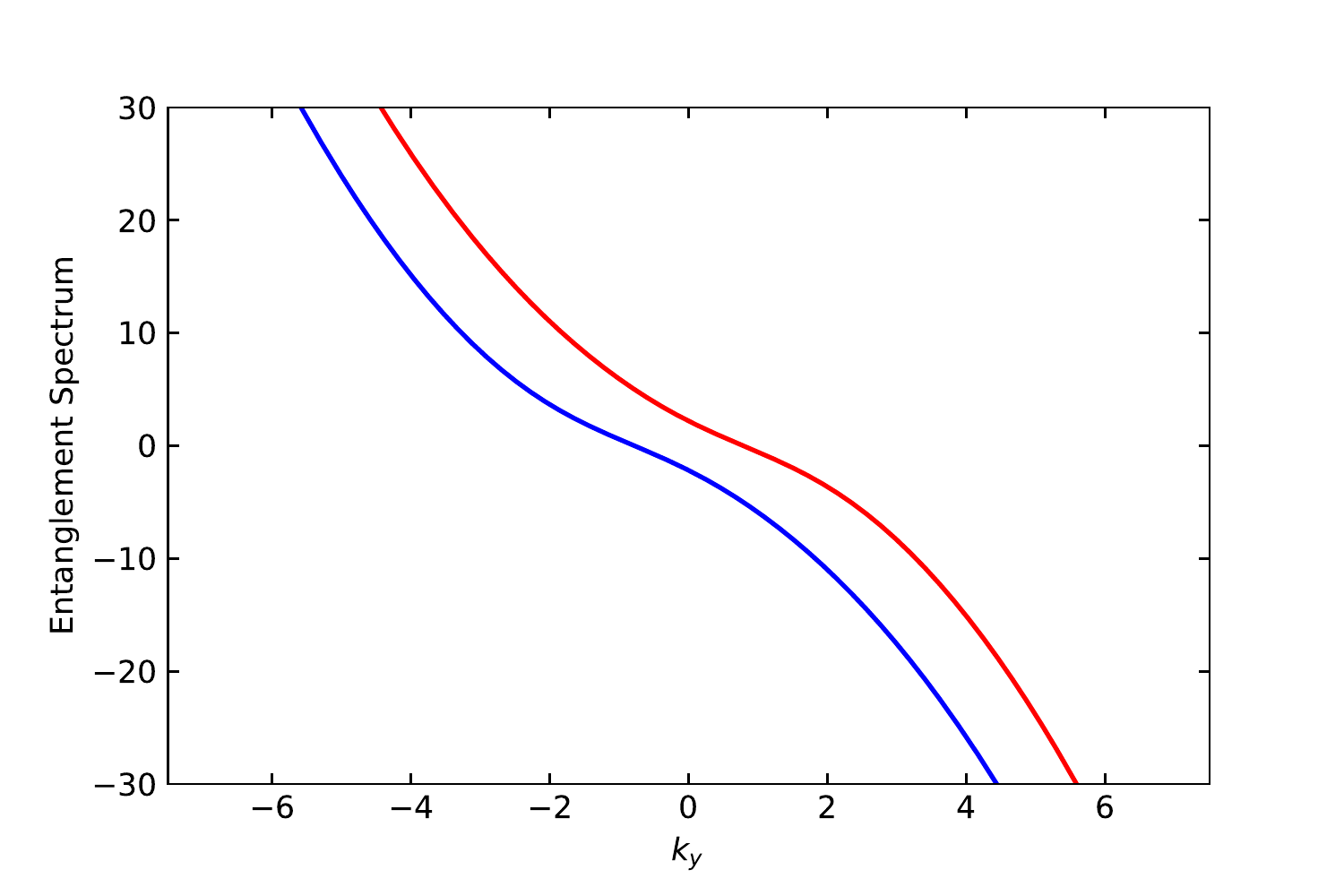}}
\subfigure[\label{spectre12ns} Arrow-shaped subregion with $\theta=\pi/2$.]{\includegraphics[width = 0.45\textwidth]{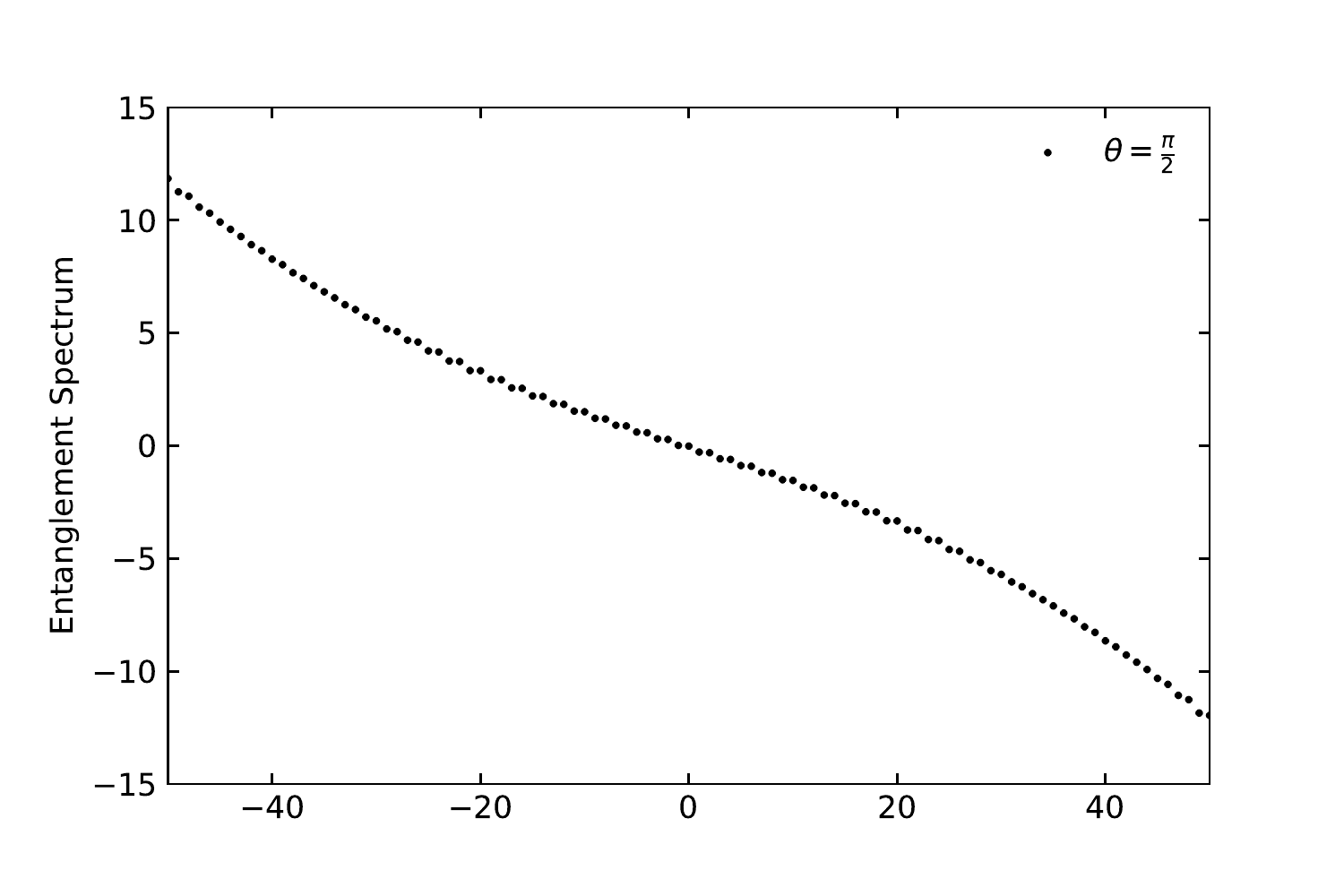}}
\caption{Entanglement spectra in the $\nu = 2$ groundstate.} 
\end{figure}

For the $\nu = 2$ state, it is harder to determine the influence of corners directly on the entanglement spectrum.
The reason is that in the smooth case, there are actually two eigenvalues for each $k$, as shown in Fig.\ts\ref{spectre12}. After computing the eigenvalues numerically for the case of the ``arrow-shaped''
domain with $\theta = \pi/2$, we obtain couples of roughly the same value at low pseudo-energy, which we guess would belong to deformations of the two distinct smooth spectra. The degeneracy is broken as the pseudo-energies move away from zero.
This situation is shown in Fig.\ts\ref{spectre12ns}.  

\subsection{Entanglement excitations} 
\begin{figure} 
\centering
\includegraphics[scale=0.85]{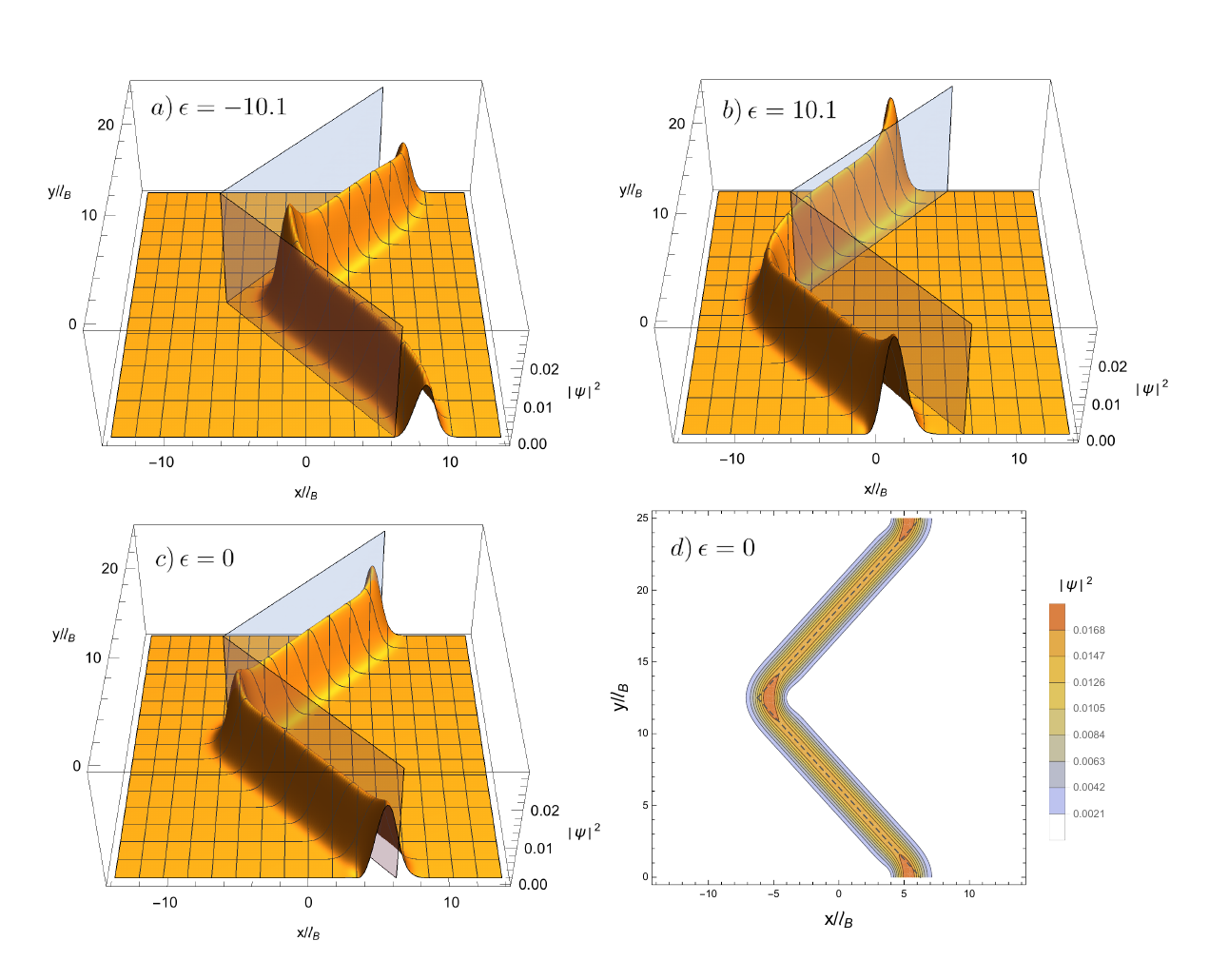}
\caption{\label{wf} Eigenfunctions of the correlator $C$. The light grey region (dotted line for the contour plot) in each panel represents the entanglement cut between $A$ and $A^c$.
  The $\epsilon = 0$ eigenfunction has zero pseudo-energy to machine precision. }  
\end{figure}

In addition to the eigenvalues of the correlator $C_{\b r,\b r'}$, one can study its eigenfunctions. Let us focus on the $\nu = 1$ ground state for brevity.
One can reconstruct a (continuous) eigenfunction of $C$, for an eigenvalue $\lambda$, from the associated discrete eigenvector of the $\mathcal{F}$ matrix as follows~\cite{SierraPolygonal}:  
${\psi (x,y) = \sum_{k} \phi_{0,k}^{*}(x,y) A_{k}}$, with $A_{k}$ the components of the normalized eigenvector associated with the eigenvalue $\lambda$. We find that the eigenfunctions associated with the low-lying part of the eigenspectrum $\epsilon$ localize close to the entanglement cut, as shown in Fig.~\ref{wf}. They decay exponentially fast 
at beyond a few magnetic lengths of the cut, as expected. 
We observe that the zero pseudo-energy eigenfunction has equal maxima at both corners, as shown in Fig.~\ref{wf} $c)$ and $d)$. 
The further the pseudo-energies $\epsilon$ are from zero, the further the eigenfunctions are located from the cut. We observe a gradual disappearance of one or the other maximum when transitioning from $\epsilon = 0$
to $\abs{\epsilon} > 0$. For the computations, we used $L_y = 25$ and $\theta = \frac{\pi}{2}$. Even though the eigenfunctions of $C$ are defined only on $A$, we have decided to show their extensions to the whole space $A \cup A^{c}$. Not surprisingly, if we had chosen the left side of the cut as $A$, we would have found that the new eigenfunctions of energy $\epsilon$ would be our old eigenfunction with energy $-\epsilon$.

We mention that similar behavior was observed for the eigenfunctions obtained in a geometry where the corner is adjacent to a physial boundary~\cite{Rozon2019}.    

\subsection{von Neumann entanglement entropy}\label{vonNeumann}
\begin{figure}
\centering
\includegraphics[scale=0.8]{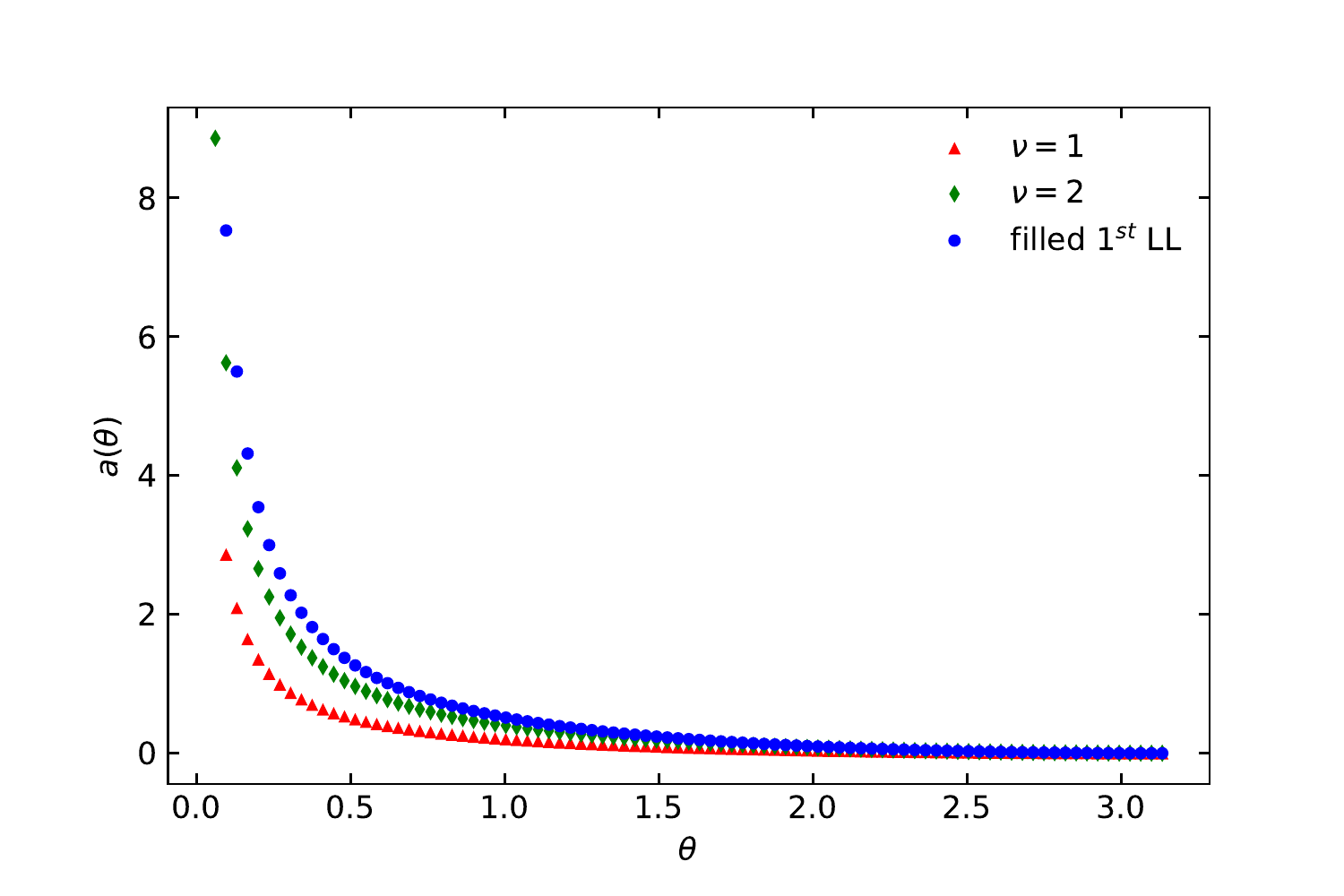}
\caption{\label{corner_fct} Corner functions for our three IQH states as a function of the opening angle $\theta$. Each curve behaves as $1/\theta$ at small angles, and $\left(\theta-\pi\right)^2$ in the smooth limit.}
\end{figure}

We can compute the von Neumann EE with the following relation
\begin{align}
  S(A)=\sum\limits_{\lambda}\left[-\lambda\ln(\lambda)-\left(1-\lambda\right)\ln(1-\lambda)\right].
\end{align}
For pure states, since $S(A)=S(A^c)$, the corner function has reflection symmetry about $\theta = \pi$: $a(\theta)=a(2\pi-\theta)$. The two corners of our arrow-shaped region thus contribute $a(\theta)+a(2\pi-\theta)=2a(\theta)$.
Also, the smooth limit gives $a(\pi)=0$ by definition since
the corner is absent in this limit. We can use this fact to determine the proportionality constant $c$ of the area law in Eq.\ts(\ref{eq:arealaw})~\cite{SierraLisse}. 
The corner function for the von Neumann EE is then obtained for our arrow-shaped domain $A$ by subtracting the area law:
\begin{align} \label{eq:a}
a(\theta)=-\frac{1}{2}\left(S(A)-c L_A \right)\,,
\end{align}
where the perimeter of $A$ is $L_A=2\sqrt{\ell_x^2+(L_y/2)^2}$. Eq.\ts(\ref{eq:a}) holds in the limit $L_A\gg 1$ so that terms of $\mathcal O(1/L_A)$ are negligible.
The required size of $L_A$ is discussed in Appendix~\ref{A:prec}.      
We show the high-precision numerical results for the corner function in Fig.\ts\ref{corner_fct}
for the $\nu=1,2$ groundstates and for the filled 1st LL excited state. A subset of the $\nu=1$ data was previously obtained (to a lower precision) in \cite{SierraPolygonal}. 
Some of the numerical data used to produce this figure are presented in Table\ts\ref{tab:table1} of Appendix\ts\ref{AppendixA}.   
 
\begin{figure}
\centering
\includegraphics[scale=0.58]{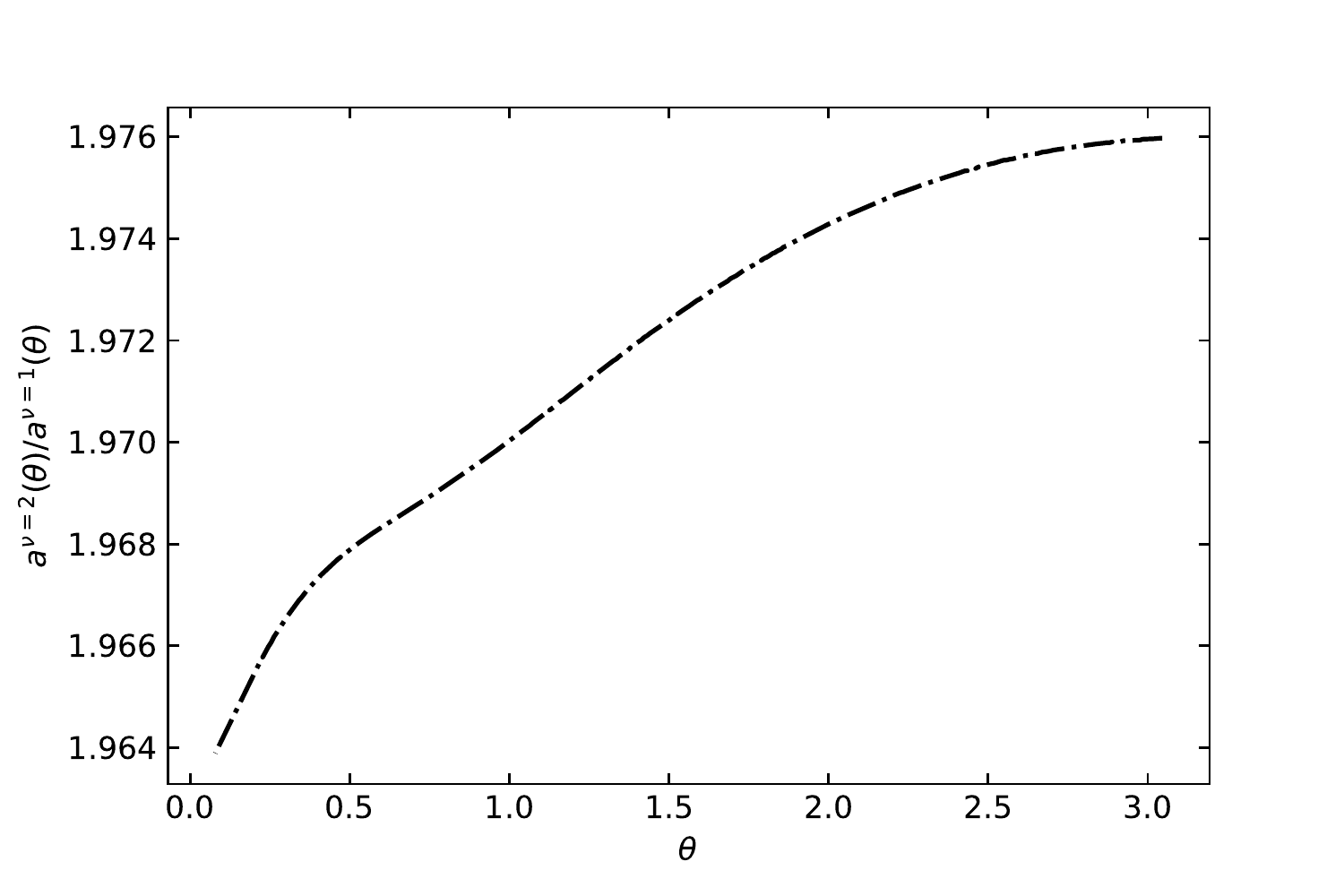}
\caption{\label{ratioCoins}Ratio of corner functions for the groundstates at $\nu=1,2$. It nearly equals the naively expected value, 2, but is clearly below it.}
\end{figure}

First, we observe that the three curves show the same small and large angle behavior, which we will discuss in more detail below.
In addition, one may naively expect that the $\nu=2$ groundstate would have a corner function that is the double of the one at $\nu=1$.  
However, the situation is not as simple since
the $n=0$ and $n=1$ wavefunctions are not only distinct, but also they are not orthogonal on subregion $A$, which leads to off-diagonal elements in the correlation matrix $\mathcal F$, Eq.\ts(\ref{eq:F-mat}).   
Surprisingly, Fig.\ts\ref{ratioCoins} shows that the ratio of the $\nu=1,2$ corner functions is almost constant and equal to 2, but is nevertheless clearly below this naive value.
In Fig.\ts\ref{corner_fct}, we also see that the 1st LL excited state has the largest corner function, even exceeding the one at $\nu=2$. This is again due to the fact that the contributions from different LLs are not additive.

In the limit of small angles, we observe the following scaling    
\begin{align} \label{kap}  
  a(\theta) = \frac{\kappa}{\theta} + \ldots
\end{align}
for all three states considered. This small angle divergence is also observed in CFTs~\cite{Casini_rev_2009,Hirata07}.
For the $\nu = 1$ state, we numerically obtain $\kappa \approx 0.276$
by analyzing the behavior of $\theta\cdot a(\theta)$ for sufficiently small angles ($\theta \approx 0.05$) 
as shown in Fig.\ts\ref{athetaRenyi}. The $\kappa$ coefficients for the three states are given in Table~\ref{tab:table2}.    

In the nearly smooth limit, we have
\begin{align} \label{smooth}
  a(\theta) = \sigma \left(\theta-\pi\right)^2 + \tilde{\sigma} \left(\theta-\pi\right)^4 + \mathcal O(\left(\theta-\pi\right)^6)
\end{align}
owing to the non-singular nature of the $\theta\to\pi$ limit, which is in contrast to the pole obtained as $\theta\to 0$, Eq.\ts(\ref{kap}).
Only even powers appear due to the reflection symmetry mentioned above, $a(\theta)=a(2\pi-\theta)$. 
For the $\nu = 1$ state, we obtain $\sigma \approx 0.02836$ and $\tilde{\sigma} \approx 0.0019$. The values of $\sigma$ for the two other states under study are given in Table~\ref{tab:table2}. 
Those results were obtained by fitting numerical values of $\frac{a(\theta)}{\left(\theta-\pi\right)^2}$ for angles near $\pi$ ($\theta \approx 3.05$) to the expected quadratic behavior.

\begin{figure}
\centering
\includegraphics[scale=0.58]{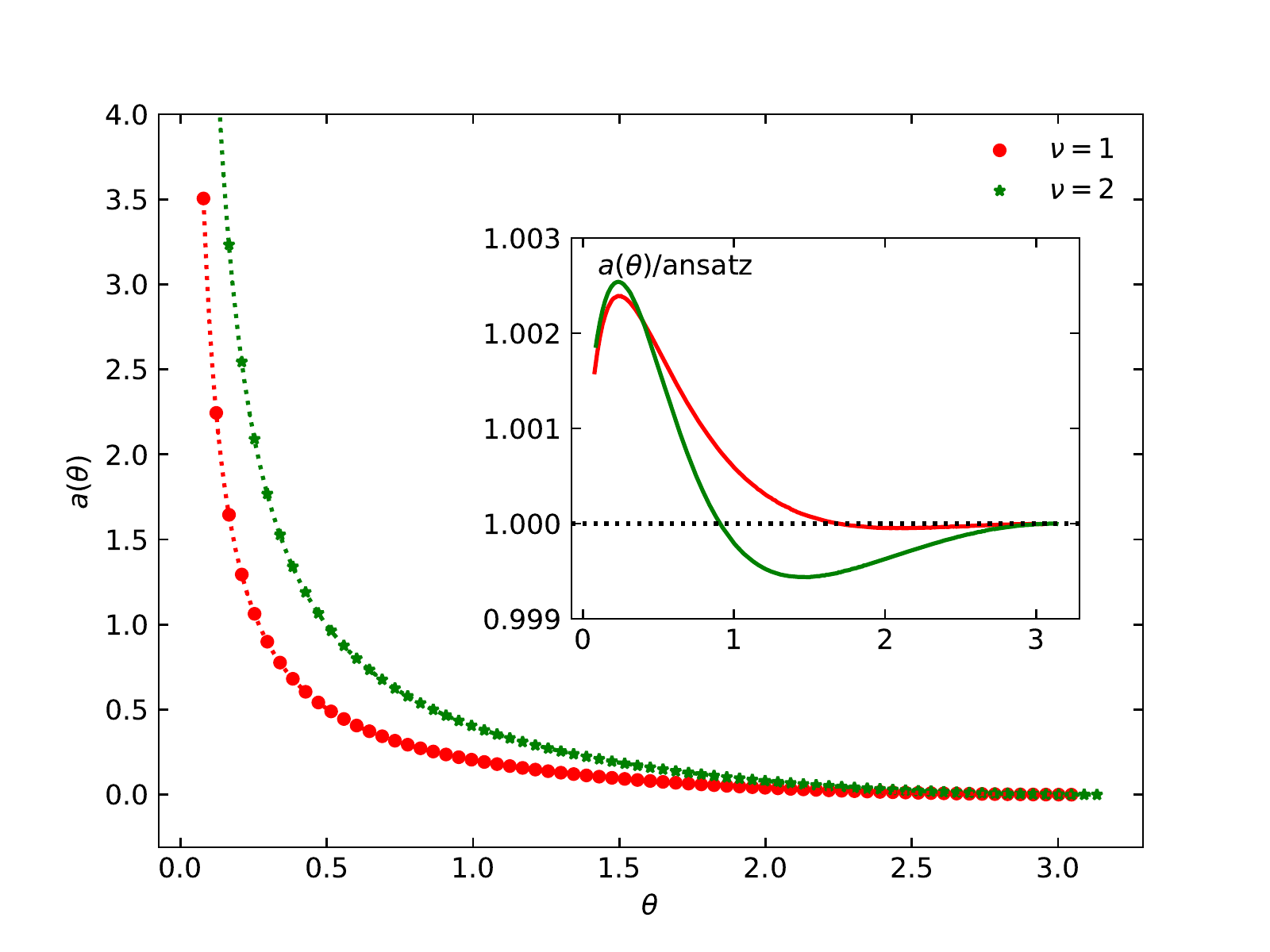}
\caption{\label{ansatz}Comparison between numerical values of the corner function for the $\nu = 1$ and $\nu = 2$ ground states to the ansatz made in \cite{BosonsFermions}. We see that the agreement between the numerical points and the ansatz, represented by the dotted lines, is excellent. The inset shows the ratio between the numerical corner function (interpolated) and the ansatz. }
\end{figure}

It is interesting to note that a simple ansatz proposed in \cite{Bueno2015} gives an approximate analytical formula for the corner function, which is exact at both asymptotic limits:
\begin{align}
a(\theta) \simeq \mu_1&\frac{(\pi-\theta)^2}{\theta(2\pi - \theta)}- \mu_2\left[1+(\pi-\theta)\cot{\theta}\right]\,,
\label{Twist}
\end{align}
where $\mu_1=2\pi\frac{\kappa - 3\pi\sigma }{\pi^2 - 6}$ and $\mu_2=\frac{3}{\pi}\frac{2\kappa - \pi^3\sigma }{\pi^2 - 6}$ are determined by the inputed smooth and sharp limit coefficients.
It is interesting to note that the function multiplied by $\mu_2$ appears in the result for the particle variance for a pie-shaped region of opening angle $\theta$~\cite{estienne2019entanglement},
although that is not the motivation behind the ansatz~\cite{Bueno2015}.    
We note in passing that this ansatz also works for the R\'enyi entropies, which will be studied in the next subsection.
From Fig.\ts\ref{ansatz}, we see that Eq.~(\ref{Twist}) works very well for both fillings. By construction,
the ratio between the ansatz and the numerical data approaches unity at small and large angles.       

The leading constant $\sigma$ in Eq.\ts(\ref{smooth}) is particularly important as we can use it to define a normalized corner function: $a(\theta)/\sigma$.
The normalized corner functions for the $\nu=1,2$ groundstates are shown in Fig.~\ref{figtan3}, where they are compared to a variety of groundstates of gapless
two-dimensional systems described by CFTs. We note that for CFTs, the corner function $a(\theta)$ comes multiplied by a logarithm $\ln(L_A/l_{\rm UV})$ owing to gapless nature of the state. This logarithm does not spoil the universality of $a(\theta)$ but only of the constant subleading term. 
We observe that the normalized corner functions for these Hall states are in fact bounded below by the one for the massless Dirac fermion CFT, and upper bounded by the massless boson CFT.  
In particular, this means that the Hall functions exceed the lower bound that holds for all CFTs~\cite{Bueno2016},    
\begin{align}
  a(\theta)/\sigma \geq 8\ln\left( 1/\sin(\theta/2) \right)
\end{align}
although the Hall Hamiltonian has a priori nothing to do with a CFT in two spatial dimensions. A stronger bound  was conjectured to hold for CFTs in~\cite{BuenoPRL}: $a(\theta)/\sigma$ is minimal for a
strongly coupled supersymmetric CFT that is holographically dual, via the AdS/CFT correspondence of string theory, to Einstein gravity in one higher dimension. In Fig.~\ref{figtan3}, it can indeed be seen that the $\nu=1,2$ curves are above the holographic one.
These findings suggests that the conjectured bound of \cite{BuenoPRL} could extend to a much broader class of quantum systems. In particular, the bound holds for the 1st LL excited state (not shown in Fig.~\ref{figtan3}) 
since $a(\theta)/\sigma$ for that state exceeds the normalized corner functions of both the $\nu=1,2$ groundstates.   

For the Hall states under study, it would be of interest to determine the physical meaning of $\sigma$.
For CFTs, $\sigma$ has in fact a very simple interpretation~\cite{BuenoPRL,Faulkner2016}: $\sigma=\pi^2 C_T/24$, where $C_T$ is the stress-tensor ``central charge''. In other words,
$C_T$ determines the 2-point function of the stress tensor (a local operator) in the groundstate of the CFT, which includes the auto-correlations of the energy density.
It would be interesting to see whether $\sigma$ for the Hall systems also possesses an interpretation in terms of local observables.

 
\subsection{R\'enyi entropies}\label{renyi EE}   
\begin{figure}
\centering
\subfigure[\label{renyi1} $\nu = 1$ state.]{\includegraphics[width = 0.48\textwidth]{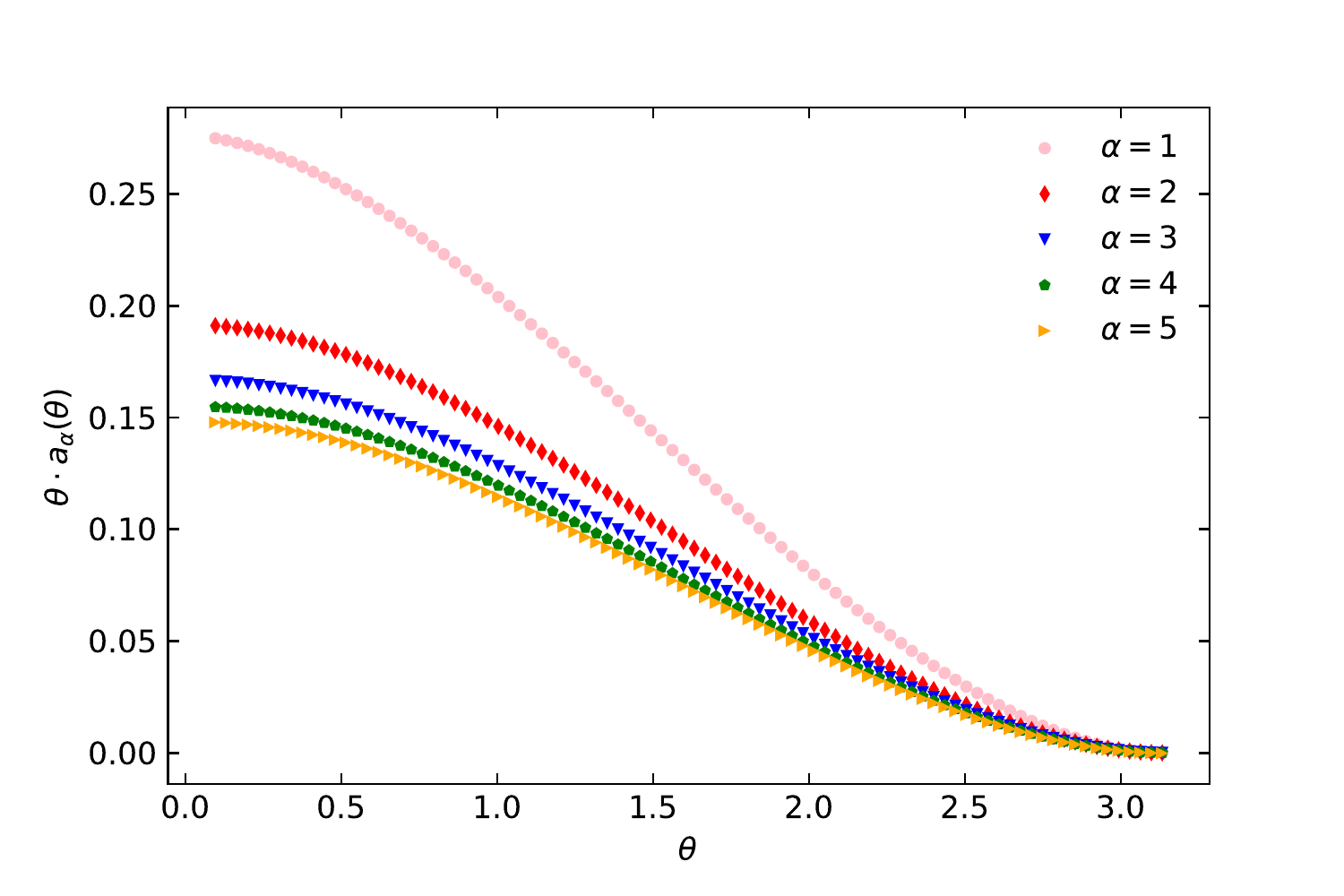}}
\subfigure[\label{renyi2} $\nu = 2$ state.]{\includegraphics[width = 0.48\textwidth]{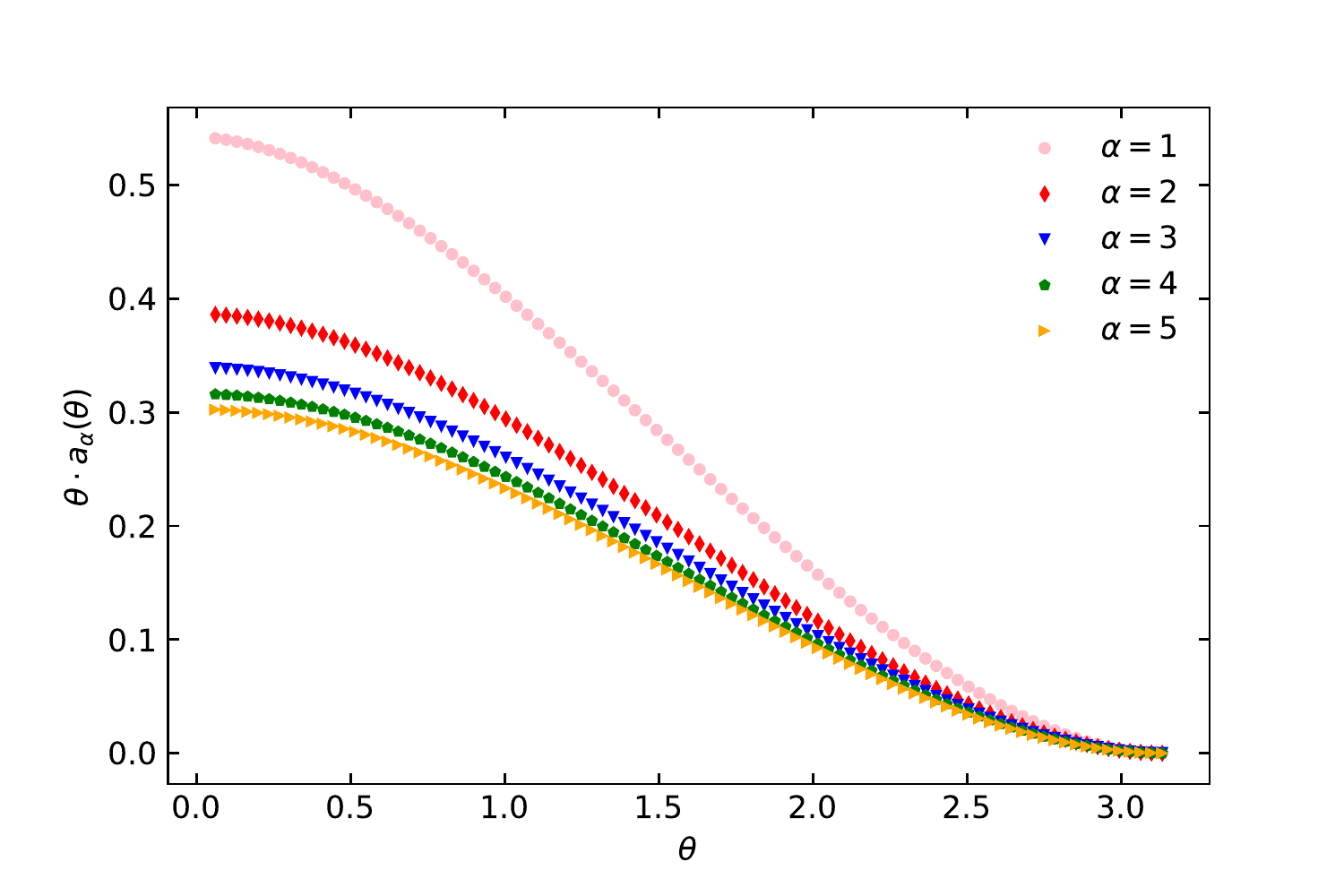}}\\
\subfigure[\label{renyi1LL} Filled $1^{st}$ LL state.]{\includegraphics[width = 0.48\textwidth]{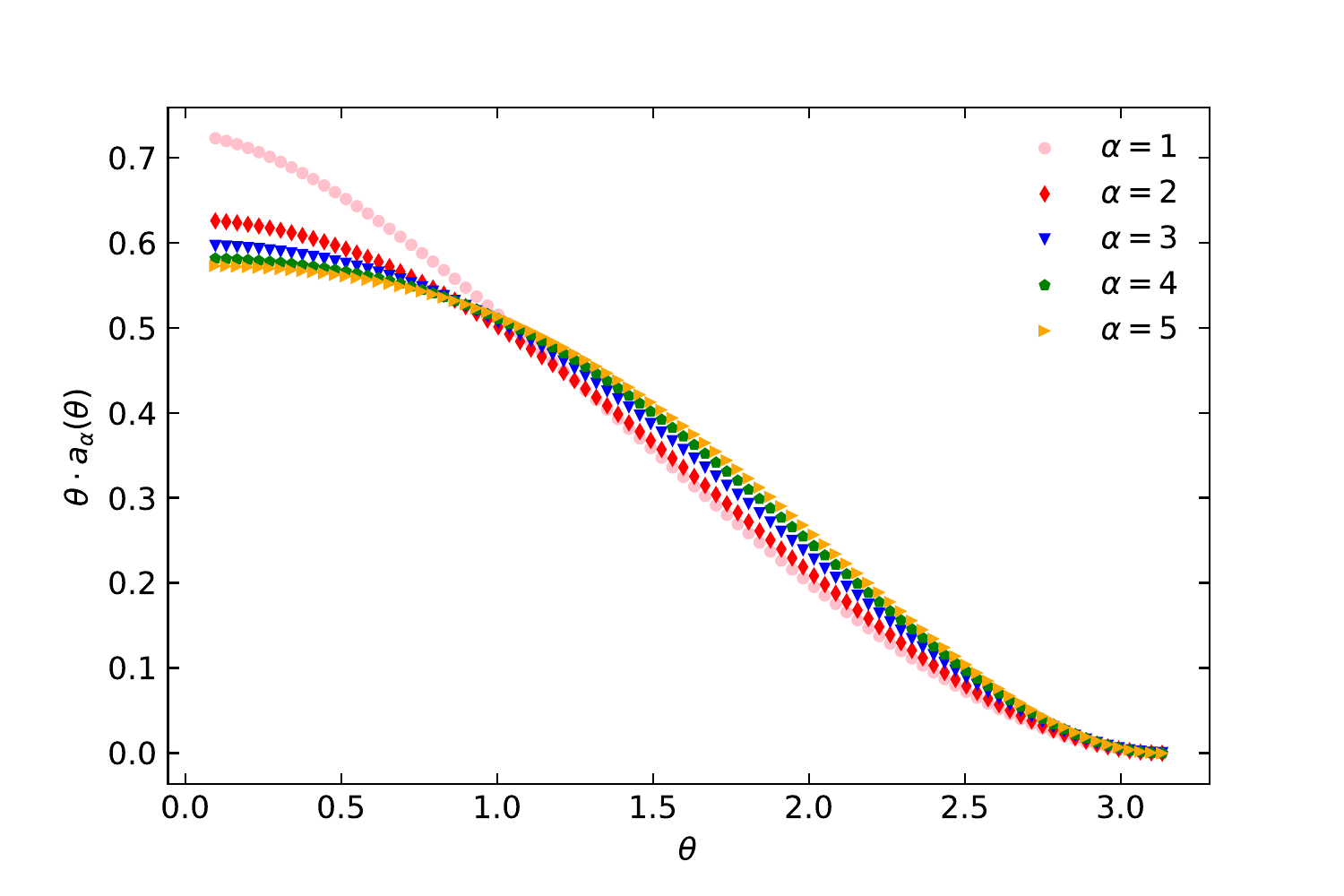}}
\caption{\label{athetaRenyi}$\theta \cdot a_{\alpha}(\theta)$ as a function of corner angle $\theta$ for the first five integer R\'enyi indices.}
\end{figure}

We now study the R\'enyi EE $S_{\alpha}(A)$ of  our IQH states, where $\alpha$ is the R\'enyi index. $S_\alpha(A)$ obeys the following large perimeter expansion: 
\begin{align}
  S_{\alpha}(A) = c_{\alpha}\frac{L_A}{\ell_B} - \sum_i a_{\alpha}(\theta_i)  + \mc O(\ell_B/L_A) 
\label{eq:perimeterLawRenyi}
\end{align}
where we have temporarily reinstated $\ell_B$.
We can then extract the corner functions by the same method as above except now EEs are connected to ESs by $S_{\alpha}(A)=\frac{1}{1-\alpha}\sum\limits_{\lambda}\ln\left[\lambda^{\alpha}+\left(1-\lambda\right)^{\alpha}\right]$.
Fig.\ts\ref{athetaRenyi} presents the corner functions for the first few integer R\'enyi indices, for each of our IQH states. 

One property of the R\'enyi EE is that it is decreasing as a function of $\alpha$. We indeed numerically verified this to be true when considering our ``arrow-shaped'' domain with two corners for multiple angles, regardless of the fact that the corner function is \textit{decreasing} as a function of $\alpha$ for the $\nu = 1$ and $\nu = 2$ states, or that it exhibits a more complicated behavior for the filled $1^{st}$ LL state. In fact, the decrease in the proportionality constant of the boundary law that is the dominant term in the studied regime where $L_A \gg 1$. It is worth noting that the non-monotonously decreasing behavior of the corner function of the excited 1st LL state is clearly distinct from
what is obtained for the groundstates at $\nu=1,2$, as well as for the massless bosons and Dirac fermion CFTs~\cite{BosonsFermions}. 

The same small and large angle behavior as described in Section~\ref{vonNeumann} is observed for the R\'enyi
corner functions $a_{\alpha}(\theta)$. The various coefficients for the asymptotic behavior are, in this case, denoted by $\kappa_{\alpha}$ and $\sigma_{\alpha}$, and their numerical values are presented in Table\ts\ref{tab:table2} for $\alpha = 2$. 

\begin{figure}[H]
\centering
\includegraphics[scale=.6]{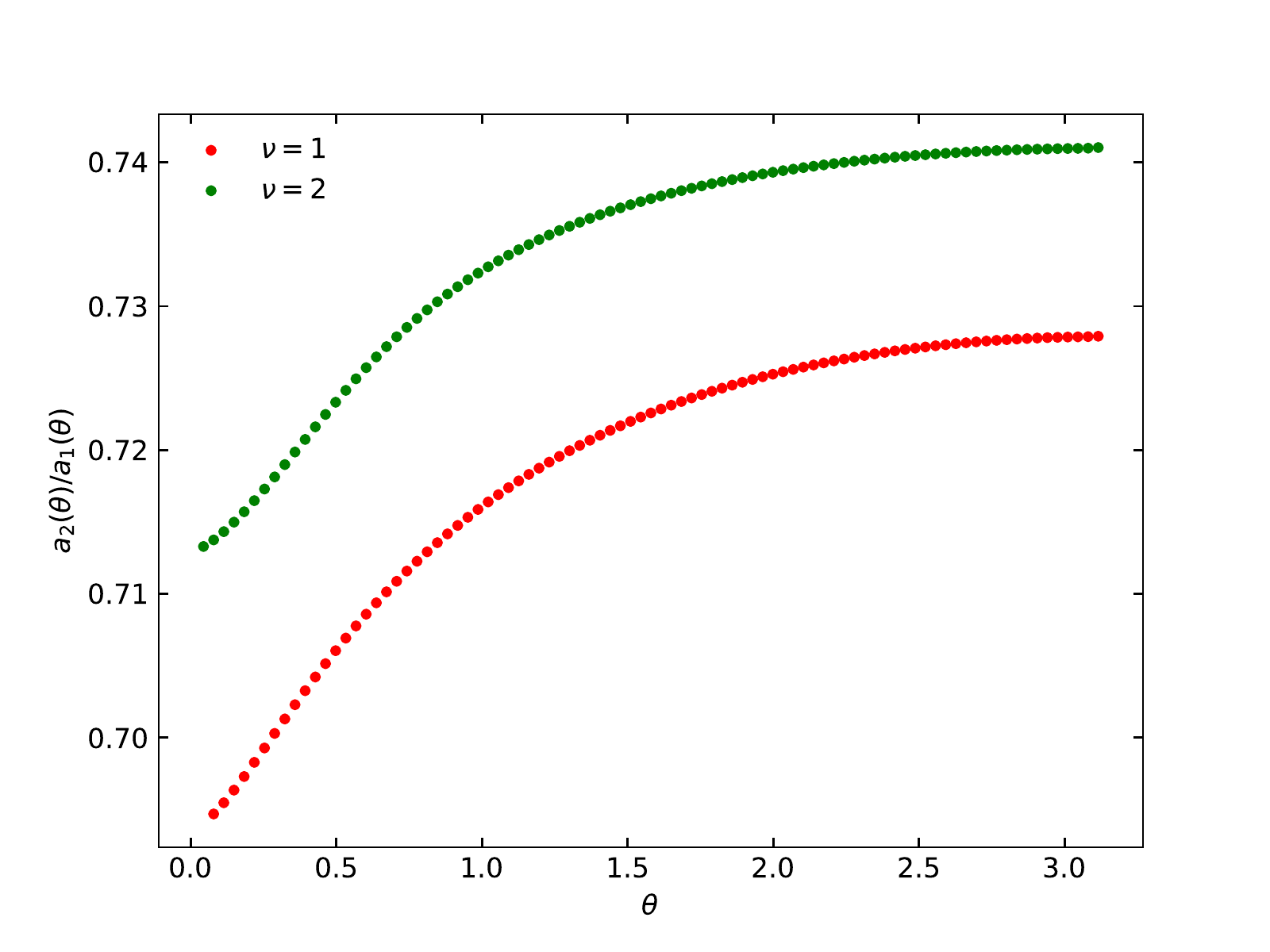}
\caption{\label{calpha}Comparison of R\'enyi EE for indices $\alpha=1,2$ for the two groundstates at fillings $\nu=1,2$.}
\end{figure}

We find that for the IQH states, the EEs of different R\'enyi indices do not factorise, i.e.\ cannot be written as $a_{\alpha}(\theta)=f(\alpha)\cdot a_1(\theta)$. Indeed, if this were the case, the ratio $a_2(\theta)/a_1(\theta)=f(2)$ would be a constant for all $\theta$, which, as shown in Fig.\ts\ref{calpha}, is not the case for $\nu = 1$ and $\nu = 2$ states. Interestingly, the observed values for the ratio $a_2/a_1$ are relatively close to what one would obtain if the R\'enyi index dependence factorized with $f(\alpha)=(1+\alpha^{-1})/2$, which is reminiscent of one-dimensional CFTs~\cite{CC2004}.    

\begin{table}[h]
  \begin{center}
    \caption{Coefficients for the asymptotic behavior of $a_{\alpha}(\theta)$ for R\'enyi indices $\alpha=1,2$.  
      Numerical values presented for $\sigma_{\alpha}$ are rounded, and were stable up to their last digit.
      The same goes for values presented in Table\ts\ref{tab:table1} of Appendix\ts\ref{AppendixA}, and in the end of Section\ts\ref{secC}. The error on numerical values of all $\kappa_{\alpha}$ is estimated at $ \pm 0.001$. }
    \label{tab:table2}
    \begin{tabularx}{\textwidth}{l*{3}{>{\centering\arraybackslash}X}*{3}{>{\centering\arraybackslash}X}}
    \toprule
    & \multicolumn{3}{c|}{$\alpha = 1$} & \multicolumn{3}{|c}{$\alpha = 2$}  \\
    & $\nu = 1$ & filled $1^{st}$ LL & \multicolumn{1}{c|}{$\nu = 2$ }& \multicolumn{1}{|c}{$\nu = 1$ }& filled $1^{st}$ LL & $\nu = 2$ \\
    \colrule
    $\sigma_{\alpha}$ & 0.02836 & 0.06895 & 0.05603 & 0.02064 & 0.07614 & 0.04152 \\
    $\kappa_{\alpha}$ & 0.276 & 0.727 & 0.542 & 0.192 & 0.627 & 0.387 \\
    \botrule
    \end{tabularx}
  \end{center}
\end{table}

\section{Anisotropic states}\label{secC}
\begin{figure}
\centering
\includegraphics[scale=0.55]{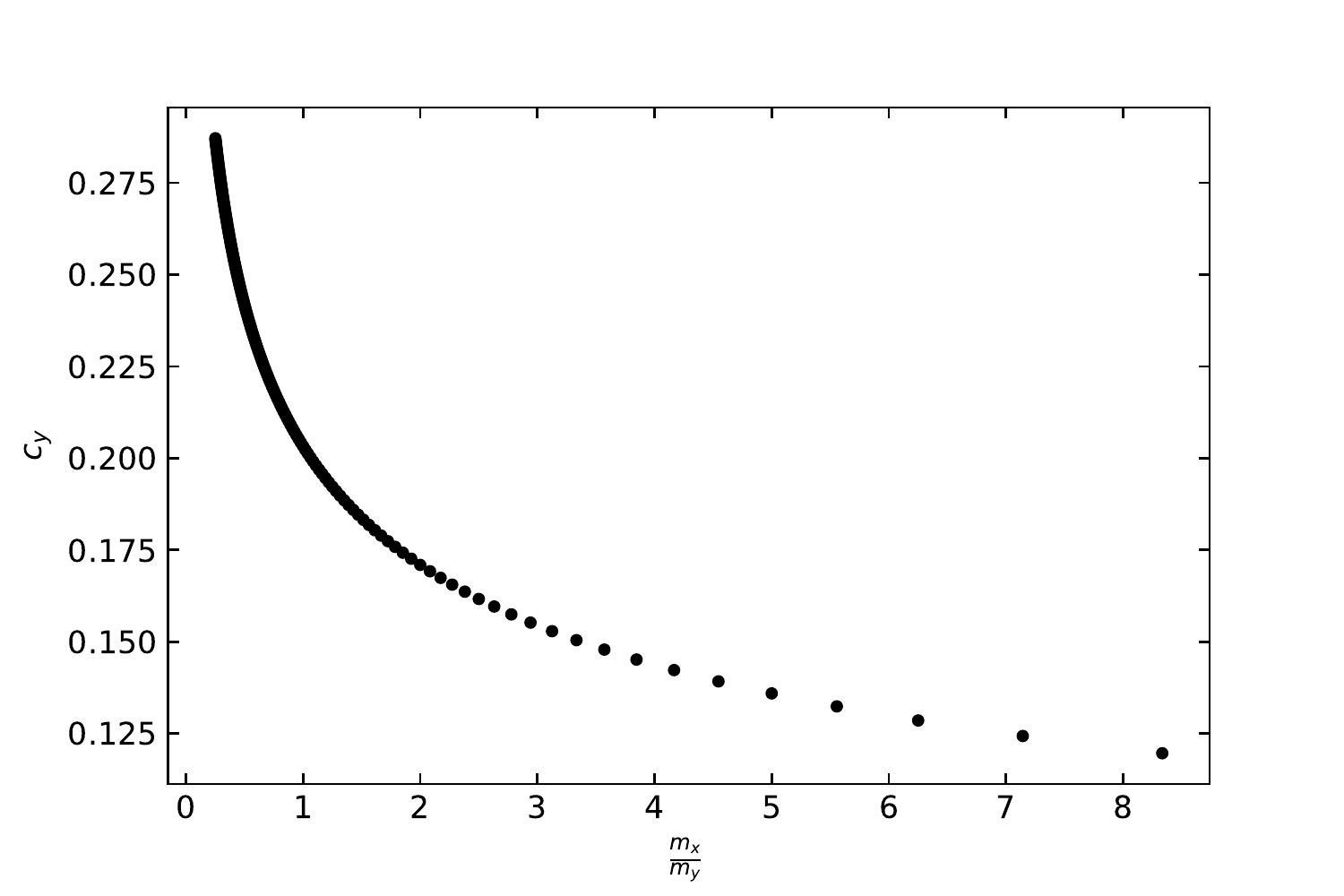}
\caption{\label{c_anis} Boundary-law coefficient for a smooth cut in the $y$ direction of the anisotropic quantum Hall state at filling $\nu = 1$ as a function of mass ratio $m_x/m_y$.} 
\end{figure}

In this section, we will study the effects of anisotropy on the EE.    
Specifically, we will break the rotational symmetry of the quantum Hall system by choosing different masses along the $x$- and $y$-directions, which results in the following
single-particle Hamiltonian:
\begin{align} 
  H=\frac{p_{x}^{2}}{2m_{x}}+\frac{\left(p_{y}+eBx\right)^{2}}{2m_{y}}\,. 
\label{h_anis}
\end{align}   
Such mass anisotropy is relevant for the description of 2DEGs with anisotropic band masses (such as AlAs or Si), uniaxial stress, or a tilted magnetic field. A more
detailed discussion about this can be found in Ref.~\onlinecite{Yang2012}.     
The anisotropic single-electron wavefunction in the LLL becomes
\begin{align}
\phi_{0,k} = \frac{1}{\pi^{\frac{1}{4}}\sqrt{\ell_BL_{y}}} \left(\frac{m_{x}}{m_{y}}\right)^{\!\frac{1}{8}} e^{ik y}  \exp\left(-\sqrt{\frac{m_{x}}{m_{y}}}\frac{\left(x+k \ell_B^{2}\right)^{2}}{2\ell_B^{2}}\right)\,,
\label{gs_anis}
\end{align} 
In the presence of anisotropy, the LL energy spectrum becomes $E_n = \hbar \tilde{\omega}_c\left(n+\frac{1}{2}\right)$, with the modified cyclotron frequency $\tilde{\omega}_c = \frac{eB}{\sqrt{m_{x}m_{y}}}$. For a smooth cut in the $y$ direction, we verify the boundary law of the von Neumann EE and obtain the proportionality constant $c_y$ ($y$ to indicate a smooth cut along the $y$ direction) for different mass ratios $m_x/m_y$.
As shown in Fig.\ts\ref{c_anis}, $c_{y}$ approches 0 as the mass ratio $m_x/m_y$ tends to infinity, which is a consequence of the fact that
electrons become more localized along the $x$-direction, see Eq.~(\ref{gs_anis}). The opposite phenomenon occurs at small mass ratio. 

\begin{figure}
\centering 

\subfigure[\label{a_anis} Corner functions (multiplied by $\theta$) for different mass ratios.
]{\includegraphics[width = 0.48\textwidth]{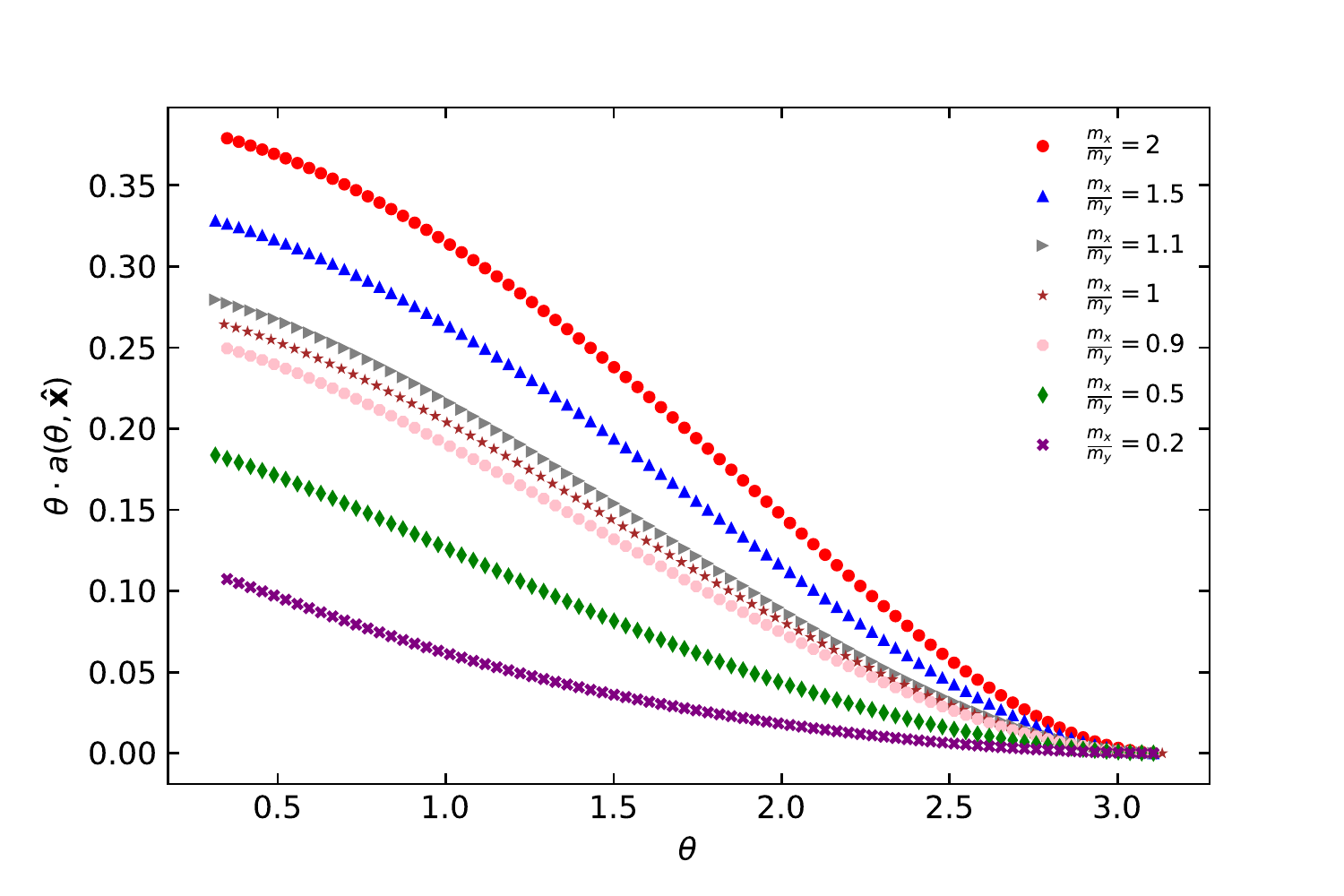}}
\subfigure[\label{theta_anis} Dependence of the corner function on the mass ratio $m_x/m_y$ for four different angles.]{\includegraphics[width = 0.48\textwidth]{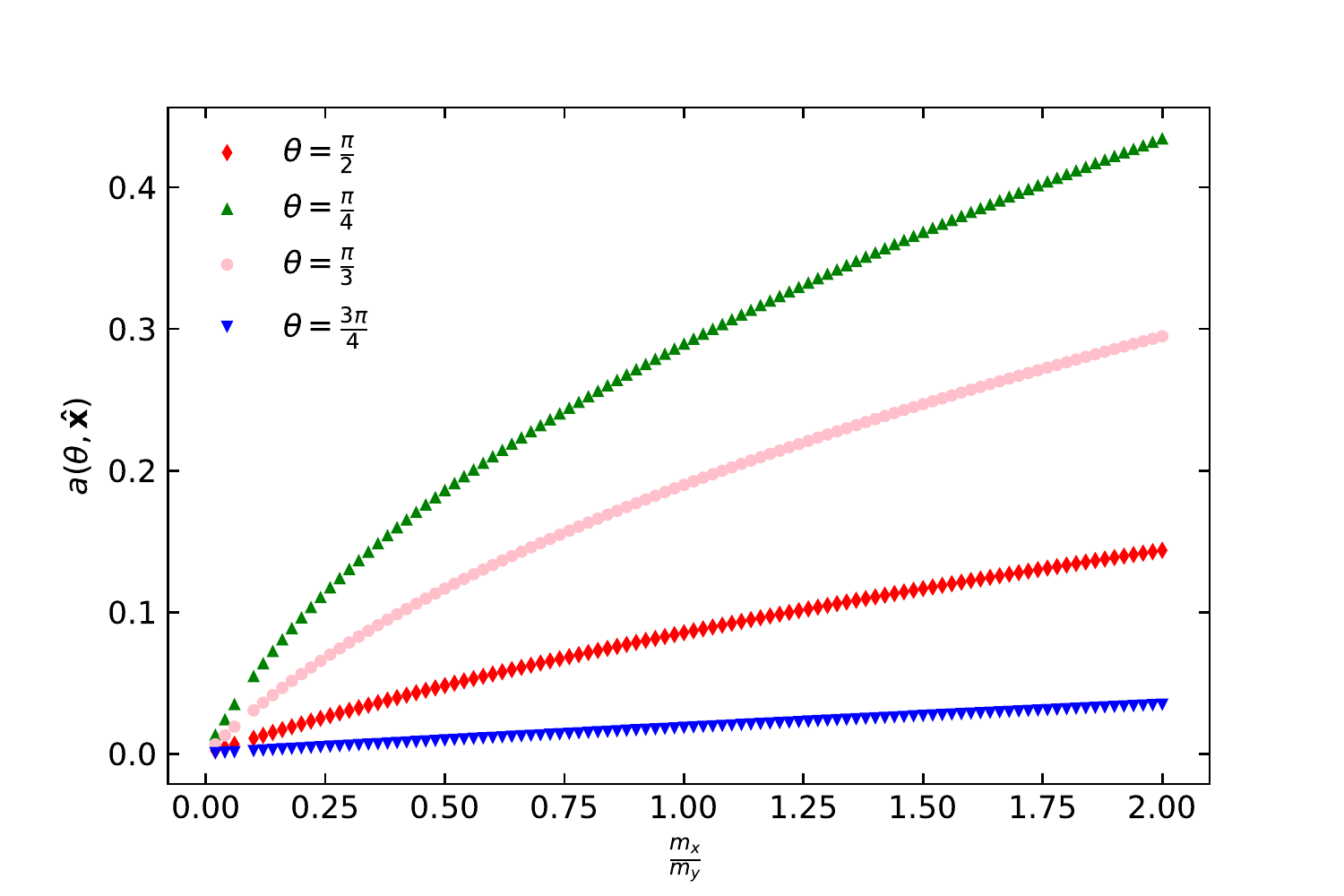}}

\caption{Mass ratio $m_x/m_y$ effect on the corner function, for corner with bisector along the $x$ axis, in the anisotropic quantum Hall state of filling $\nu = 1$
  for the ``arrow-shaped'' domain.}
\end{figure}

For our arrow-head geometry, the breaking of rotational symmetry modifies the boundary law as follows:
\begin{align}
2c \sqrt{\ell_x^2+\left(L_y/2\right)^2} \to 2\sqrt{\left(c_{x}\ell_{x}\right)^{2}+\left(c_{y} L_{y}/2 \right)^{2}} \label{2eq29} \,,
\end{align}
where $c_x$ is the boundary-law constant that would be obtained for a flat cut parallel to the $x$-axis.
  By symmetry, $c_x$ at a given mass ratio is given by $c_y$ with the ratio inverted, $c_x(m_x/m_y)=c_y(m_y/m_x)$.
We note that $c_x=c_y$ only if the masses are equal.  
The corner contributions also inherit such a dependence on orientation.
In the presence of mass anisotropy, we denote the corner function $a(\theta,\boldsymbol{\hat{u}})$, where $\boldsymbol{\hat{u}}$ is the unit vector parallel to the bisector of the corner $\theta$ pointing inwards of subregion $A$.
By assuming that the contribution of a corner is unchanged by a reflection of this corner along the $x$ and/or $y$ axis,
which seems reasonable considering the symmetries of the system, we should still be able to extract the contribution of a corner with bisector oriented along the $x$ axis $a(\theta,\boldsymbol{\hat{x}})$ from the same ``arrow-shaped'' region of Fig.\ts\ref{schema} used until now. We can then study the dependence on the mass ratio $m_x/m_y$ of a corner of this particular orientation,  
which is what is presented in Fig.\ts\ref{a_anis} for the $\nu=1$ state. 
From Fig.\ts\ref{theta_anis}, we see that the corner function for a given opening angle grows with the mass ratio $m_x/m_y$.
This growth leads to a \emph{decrease} of the total EE since the corner contribution appears with a negative sign, $-\sum_i a(\theta_i,\boldsymbol{\hat{u}}_i)$.

\begin{figure}
\centering
\includegraphics[scale=0.7]{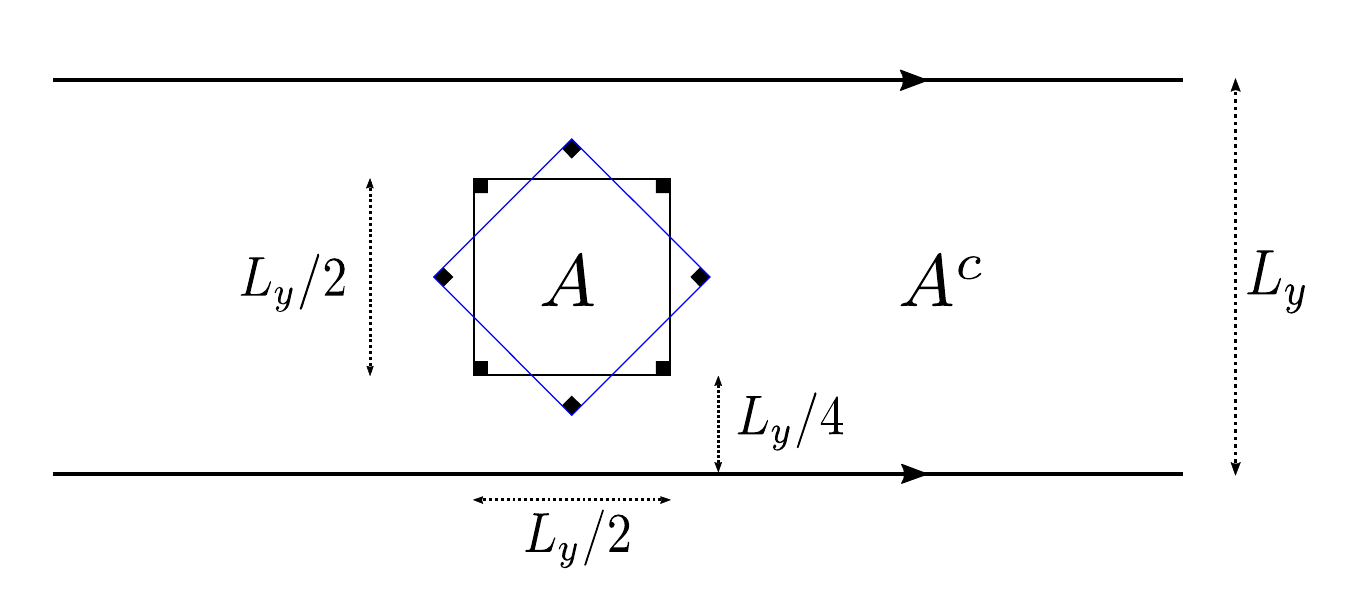}  
\caption{\label{schema2}Square subregions $A$ and their complement on an infinite cylinder of circumference $L_y$: the black square is aligned with
    the $x$ and $y$ axes, while the blue one is rotated by $\pi/4$. In the presence of mass anisotropy, they give different entanglement entropies.} 
\end{figure}

Finally, we observe a peculiar behavior of the orientation-dependence of the corner function. 
By computing the EE for a simple square with edges parallel to the $x$ and $y$ axes (see Fig.\ts\ref{schema2}), we observe \emph{no} dependence of the corner function on the mass ratio, so that the corner function (tested for mass ratios $m_x/m_y \in \{2,5,50\}$) corresponds to that for a unit mass ratio to at least to 8 significant digits: $a(\frac{\pi}{2},\frac{1}{\sqrt{2}}\left(\boldsymbol{\hat{x}}+\boldsymbol{\hat{y}}\right))=0.085498696$. We also find by
considering a rotated square with angle bisectors oriented along the $x$ and $y$ axis (see Fig.\ts\ref{schema2}), that the corner contribution $a(\frac{\pi}{2},\boldsymbol{\hat{y}})$ is that of a corner whose bisector is oriented
along the $x$ axis $a(\frac{\pi}{2},\boldsymbol{\hat{x}})$ but with an inverted mass ratio, consistent with the $\pi/4$ rotational symmetry. 
For example, we can extract from a calculation with the arrow-shaped domain that the contribution of a corner whose bisector is oriented along the $x$ axis that for a mass ratio $m_x/m_y = \frac{1}{2}$, $a(\frac{\pi}{2},\boldsymbol{\hat{x}}) = 0.0481848$. This way, when considering our tilted square, we should have  $a(\frac{\pi}{2},\boldsymbol{\hat{y}})=-\frac{1}{2}\left(S-2\cos{\frac{\pi}{4}}\sqrt{c_x^2+c_y^2} L_y -2a(\frac{\pi}{2},\boldsymbol{\hat{x}})\right)$, i.e. the contribution from each corner with bisectors along the $y$ axis. This gives $a(\frac{\pi}{2},\boldsymbol{\hat{y}}) = 0.143702$, which is (to at least 6 significant digits) $a(\frac{\pi}{2},\boldsymbol{\hat{x}})$ for a mass ratio $m_x/m_y = 2$. 

\section{Tip-touching corners}\label{sec:tip-touch} 

\begin{figure}
\centering
\includegraphics[scale=0.7]{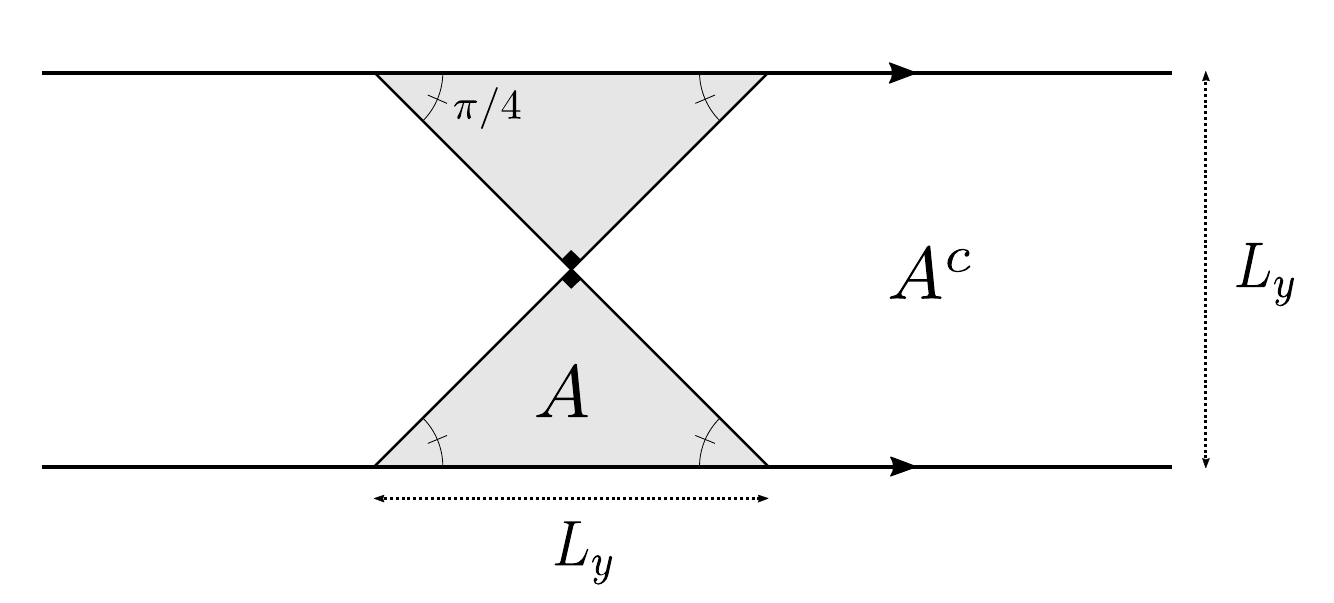}
\caption{\label{schema3} ``Hourglass'' subregion $A$ and its complement on an infinite cylinder of circumference $L_y$. Note that $A$ contains four $\pi/2$ corners, two of which touch at their vertex. }
\end{figure}

Until now, we have studied geometries for which the corner contribution to the EE is additive, see Eq.~(\ref{eq:cornercontribution}).  
One instance where this additivity fails is when two or more corners touch at their tip.
As an example, consider the von Neumann EE in the $\nu = 1$ state for the ``hourglass'' geometry in Fig.\ts\ref{schema3}. 
We find that $\gamma \neq 4a(\pi/2)$, but that the contribution from the two touching corners (the other two are simply the bulk corners studied above),
which we denote by $a_{\cross}(\pi/2)$, is 
\begin{align} \label{eq:across}
 a_{\cross}(\pi/2) = \gamma - 2a(\pi/2) = 0.379024 > 2a(\pi/2) 
\end{align}
where $2a(\pi/2)=0.170997$. In fact, $a_{\cross}(\pi/2)$ is close (but not equal) to \emph{four} times the $\pi/2$ contribution, $4a(\pi/2)$. This clearly shows the failure of the additivity for touching corners.    
We have also obtained the result for the second R\'enyi entropy, $\alpha=2$: $a_{\cross,2}(\pi/2)  = \gamma_2-2a_2(\pi/2) = 0.232144 > 2a_2 (\pi/2)$.

 For this geometry, it is of interest to compute the mutual information, $I(A_1,A_2)=S(A_1) + S(A_2) -S(A_1\cup A_2)$, where $A_{1,2}$ are two subregions.
  Here, we take $A_1$ to be the top part of the hourglass, and $A_2$ the bottom part. We can make a simplification by ignoring what happens far from the point where the two corners meet.
  This could be achieved by working with an infinite hourglass embedded in the plane, or by making the vertical extent of the hourglass smaller than $L_y$
  and smoothing out the bulk corners so that the hourglass is the only singularity. In that case, we get the following mutual information

  \begin{align}
    I(A_1,A_2)= a_{\cross}(\pi/2) - 2a(\pi/2)=0.208027
  \end{align} 
  The boundary law part cancelled out, and we are left with an expression independent of all scales. Further, by the subadditivity of the EE, we have that the
  mutual information is non-negative so that $a_{\cross}(\pi/2) \geq 2a(\pi/2)$. Our numerical result is clearly consistent with this constraint.   

\section{Conclusion}\label{sec:concl}
We have studied the properties of the reduced density matrix for various regions with sharp corners in three IQH states: two groundstates at
fillings $\nu=1,2$ and one excited state with the 1st LL entirely full (and the 0th and other LLs empty). We have studied the non-trivial angle dependence of the EE (and its R\'enyi generalizations), which is encapsulated by the corner function $a(\theta)$. This function is independent of all scales, including the magnetic length $\ell_B$. Most strikingly, we found that the Hall corner function, when properly normalized, has a surprisingly close
angle dependence to the corner function found in two-dimensional CFTs, as shown in Fig.~\ref{figtan3}. This super-universality of the EE hints at common structures in very different quantum many-body states. 
In particular, as we discussed in Section~\ref{vonNeumann}, the Hall corner terms obey bounds that hold for CFTs. It would be desirable to understand the reason for
these common properties. In this regard, quantum information theoretic concepts could reveal general properties about the entanglement structure of a large class of quantum many-body states.

We also examined the role of mass anisotropy on the EE, and showed that it strongly affects the shape dependence of the EE. 
We studied a different type of corner where two tips touch at a point (``hourglass''), and we were able to extract a universal quantity via the mutual information.

Going beyond the EE, we studied the entanglement spectrum, as well as the eigenfunctions of the reduced density matrix. In particular, we described how the eigenfunctions associated with low pseudo-energy eigenstates localize near sharp corners. The behavior of the entanglement spectrum also shows clear differences between the ground and excited states.

\subsection{Outlook}
In this work, we have shown that IQH states provide a rich playground to study the entanglement structure of quantum states. The states we have studied are particularly simple, but nevertheless share common properties with much more complicated states such as the groundstates of interacting two-dimensional CFTs. It would be of interest
to study the same quantities in more intricate topological states, such as the Laughlin FQH states.
For instance, it would be desirable to obtain the corner function $a(\theta)$ for such FQH states, and compare its angle dependence with that of IQH states, and two-dimensional CFTs. This is a challenging task owing to the non-Gaussian nature of FQH states, but one could make use of the recent advances in representing trial wavefunctions using Matrix Product States (MPS)~\cite{MPS12,MPS13}.                  
In particular, the MPS representation works for the infinite cylinder geometry used in this work.
Apart from the numerical analysis, analytical results could be obtained, in particular in the nearly smooth limit $\theta\approx\pi$, where analytical results already exist for
general CFTs~\cite{Bueno2015,Faulkner2016}.   

In the context of two-dimensional CFTs, it was recently realized that the EE of a region that intersects a physical edge of the system has a relation to the EE of a bulk region obtained by taking the union of the initial region and its mirror image about the edge~\cite{Berthiere2019}. This relation was shown to hold approximately for the IQH groundstate at $\nu=1$ for a large range of intersection angles, and exactly for a specific angle~\cite{Rozon2019}. It would be desirable to study this relation in the other IQH states studied in this work, as well as in FQH states.     

\begin{acknowledgements}  
  We thank G.~Sierra for sharing with us many results, both published and unpublished, on entanglement in quantum Hall states. We also thank C.~Berthiere, A.~Gromov, F. D. M.~Haldane, L.~Santos, P.-G.~Rozon for useful discussions.
This project was funded by a grant from Fondation Courtois, a Discovery Grant from NSERC,
a Canada Research Chair, and a ``Etablissement de nouveaux chercheurs et de nouvelles chercheuses universitaires'' grant from FRQNT. 

This work was initiated and partially performed at the Aspen Center for Physics, which is supported by National Science Foundation grant PHY-1066293.  
\end{acknowledgements}        
 
\clearpage
\appendix

\noindent{\LARGE \bf Appendices} \\

\section{Numerical values} \label{AppendixA}
We give the numerically calculated corner function $a_\alpha(\theta)$ in Table~\ref{tab:table1}, where $\alpha$ is the R\'enyi index.
According to our analysis, described in detail in Appendix~\ref{A:prec}, all the quoted digits are numerically stable.  

\begin{table}[h]
  \begin{center}
    \caption{Numerical values of $a_{\alpha}(\theta)$.}
    \label{tab:table1}
    \begin{tabularx}{\textwidth}{l*{3}{>{\centering\arraybackslash}X}*{3}{>{\centering\arraybackslash}X}}
    \toprule
    & \multicolumn{3}{|c|}{$\alpha = 1$} & \multicolumn{3}{c}{$\alpha = 2$}  \\
    $\theta$ [$^{\circ}$] & \multicolumn{1}{|c}{$\nu = 1$} & filled $1^{st}$ LL & \multicolumn{1}{c|}{$\nu = 2$} & \multicolumn{1}{c}{$\nu = 1$ }& filled $1^{st}$ LL & $\nu = 2$  \\
    \colrule
    5 & 3.15235 & 8.29181 & 6.19120 & 2.19050 & 7.17506 & 4.41980 \\ 
    10 & 1.56134 & 4.09623 & 3.06824 & 1.08834 & 3.57006 & 2.19540 \\ 
    15 & 1.02634 & 2.68343 & 2.01796 & 0.717969 & 2.36081 & 1.44788 \\ 
    25 & 0.591423 & 1.53394 & 1.16366 & 0.416631 & 1.38026 & 0.839960  \\ 
    35 & 0.399602 & 1.02778 & 0.786566 & 0.283233 & 0.947825 & 0.570982  \\ 
    45 & 0.289663 & 0.739082 & 0.570368 & 0.206366 & 0.698379 & 0.415970 \\ 
    55 & 0.217535 & 0.550975 & 0.428509 & 0.155638 & 0.532717 & 0.313651 \\ 
    65 & 0.166224 & 0.418217 & 0.327573 & 0.119344 & 0.412985 & 0.240445 \\ 
    75 & 0.127776 & 0.319567 & 0.251913 & 0.0920046 & 0.321673 & 0.185314 \\ 
    85 & 0.0979695 & 0.243721 & 0.193228 & 0.0707108 & 0.249586 & 0.142389 \\ 
    90 & 0.0854987 & 0.212181 & 0.168664 & 0.0617735 & 0.219010 & 0.124377 \\ 
    95 & 0.0743545 & 0.184105 & 0.146707 & 0.0537720 & 0.191448 & 0.108255 \\ 
    105 & 0.0554144 & 0.136645 & 0.109373 & 0.0401398 & 0.144027 & 0.0807942 \\ 
    115 & 0.0401632 & 0.0986867 & 0.0792945 & 0.0291311 & 0.105244 & 0.0586257 \\
    125 & 0.0279366 & 0.0684378 & 0.0551689 & 0.0202846 & 0.0737173 & 0.0408167 \\ 
    135 & 0.0182760 & 0.0446606 & 0.0360985 & 0.0132814 & 0.0485057 & 0.0267218 \\ 
    145 & 0.0108613 & 0.0264893 & 0.0214566 & 0.00789824& 0.0289604 & 0.0158896 \\ 
    155 & 0.00547020 & 0.0133214 & 0.0108077 & 0.00397976 & 0.0146362 & 0.00800594 \\ 
    165 & 0.00195267 & 0.00475064 & 0.00385825 & 0.00142108 & 0.00523667 & 0.00285860 \\ 
    170 & 0.000865582 & 0.00210523 & 0.00171033 & 0.000629997 & 0.00232299 & 0.00126727 \\ 
    175 & 0.000216057 & 0.000525388 & 0.000426921 & 0.000157262 & 0.000580090 & 0.000316338 \\ 
    \botrule
    \end{tabularx}
  \end{center}
\end{table}

\newpage
\section{Precision}\label{A:prec}
\begin{figure}[H]
\centering
\includegraphics[scale=.9]{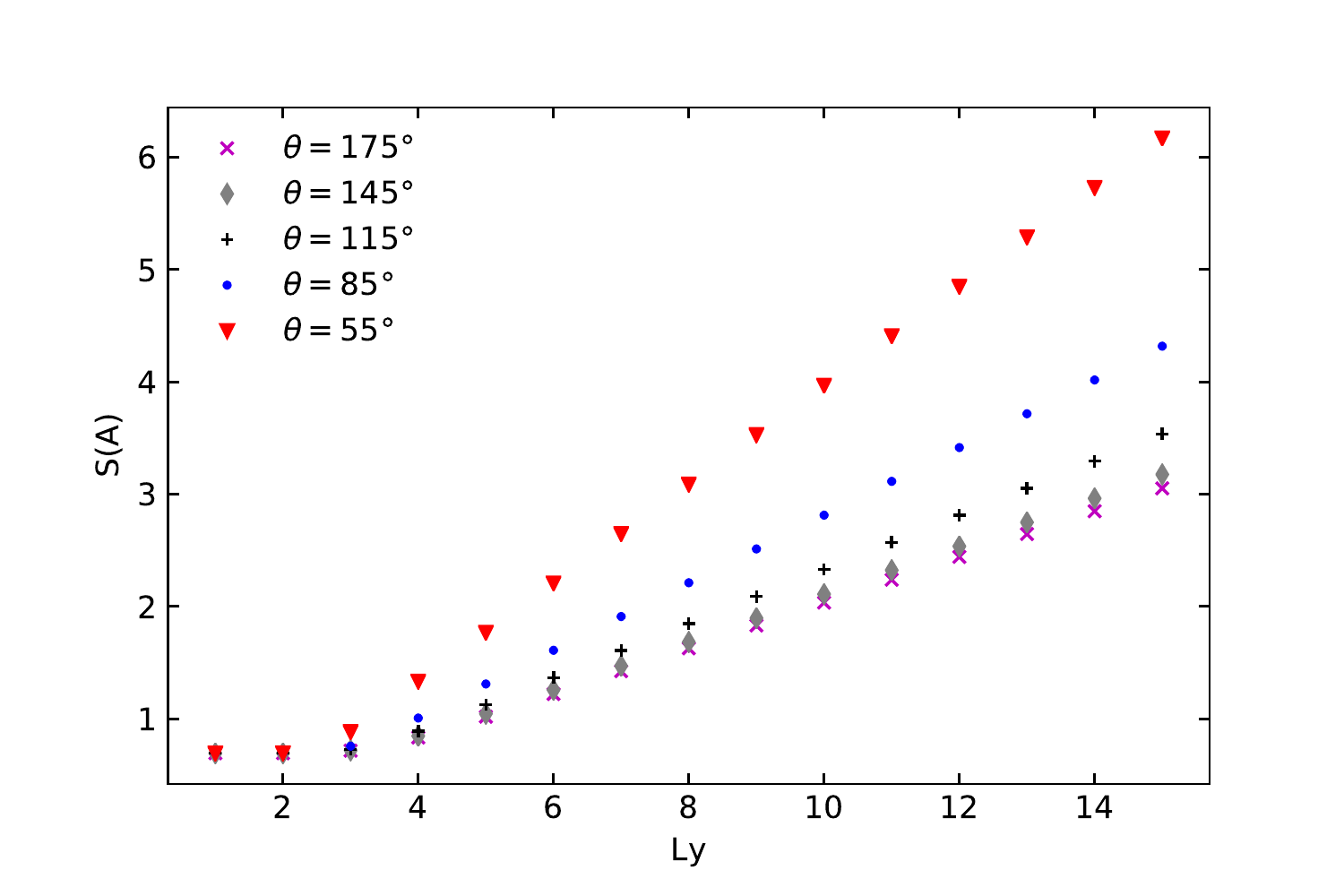}
\caption{Von Neumann EE as a function of cylinder circumference $L_{y}$ for the $\nu = 1$ groundstate. We note the transition to the boundary law regime for every ``arrow-shaped'' region of angles $\theta$. At large $L_y$, we recover the boundary law $S(A) = c \frac{L_A}{\ell_B} - 2 a(\theta) + ... = c \csc{(\frac{\theta}{2})} \frac{L_y}{\ell_B} - 2 a(\theta) + ... $}
\label{prec1}
\end{figure}

The EE results from the diagonalization of the infinite-dimensional matrix $\mathcal{F}(A)$. To perform numerical computations, we had to truncate this matrix by ignoring terms related to electrons located further
from the entanglement cut. These electrons contribute less to the EE, and are associated with high absolute value of the momentum $ k$.
In addition to the size of the matrix, it was necessary to consider, for every computation, a sufficiently large value for the circumference of the cylinder $L_{y}$. Indeed, we needed to consider a cylinder that is big enough for the area law to hold. The method implemented to ensure the convergence of the corner function of the von Neumann EE $S(A)$ for any of the three states is the following. Concretely, for a desired precision $\sim \delta$ on a value of the corner function $a(\theta)$ (for a given angle $\theta$), we first need to fix $L_y$ and compute the EE from a matrix whose dimension $N$ increases until it is stable at a precision of $\sim 2 \cdot \delta$ (2 for the number of corners in our region, this way the precision on the (possibly erroneous) corner function extracted from this EE is $\sim \delta$). Then, we redo this step for a significantly larger $L_y$. From these two EEs, the value of the corner function is calculated and compared to confirm/infirm that we are in the area law regime. These two steps are then repeated until the precision on the corner function reaches $\sim \delta$. We note that the minimum dimension $N$ for the EE convergence seems directly proportional to $L_y$ and Fig.\ts\ref{prec1} shows the transition to the boundary law regime for the von Neumann EE $S(A)$ as a function of $L_y$, for multiple ``arrow-shaped'' subregions $A$ of angles $\theta$. 

\begin{figure}[h]
\centering
\includegraphics[scale=0.79]{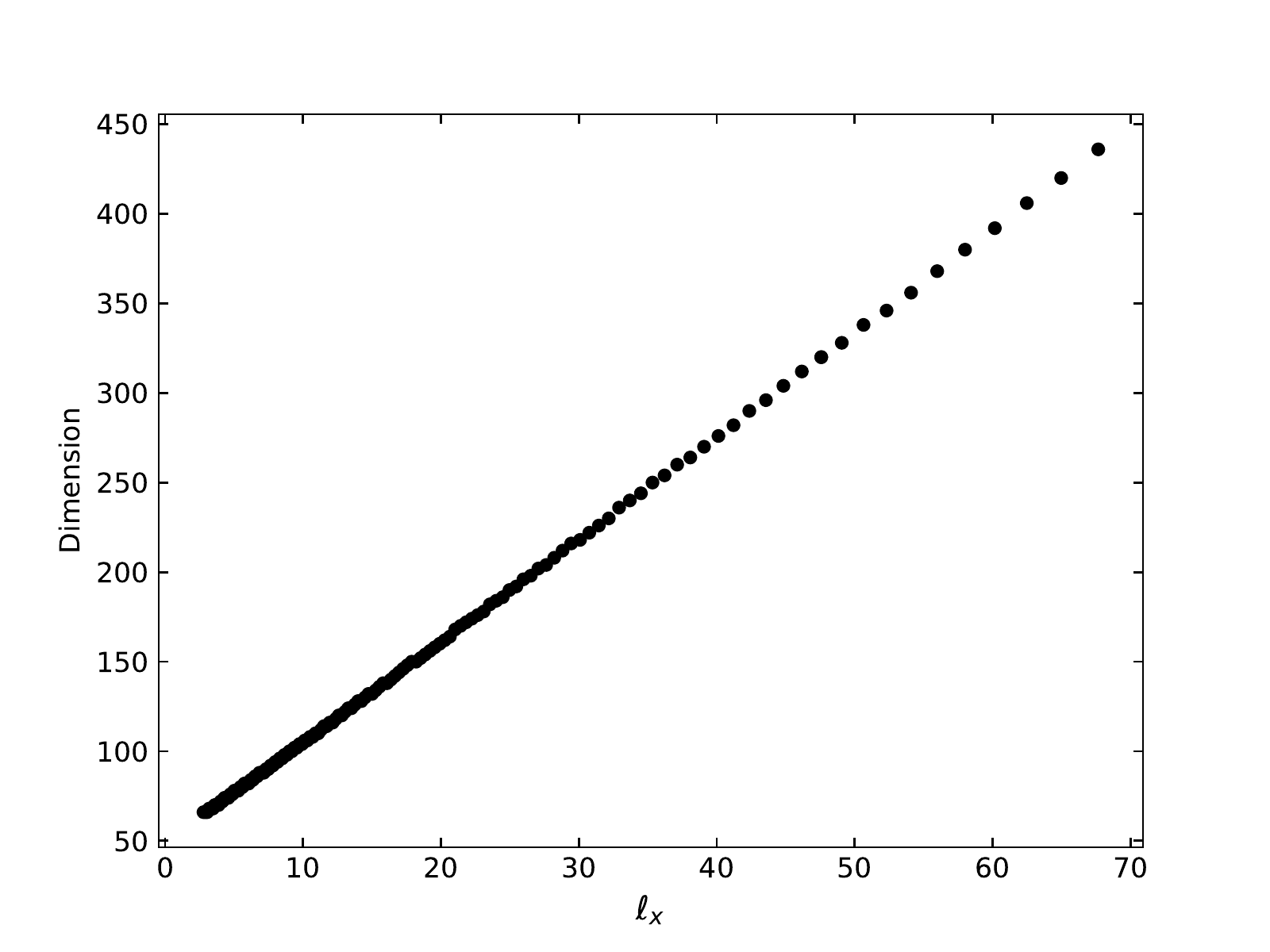}
\caption{Minimal dimension $N$ of $\mathcal{F}(A)$ (so that we consider $k$ of quantum number $m \in \left[-N/2+1,N/2\right]$ in $\mathcal{F}(A)$) as a function of $\ell_x$, for a stable $a(\theta)$ for the LLL state with a precision $10^{-8}$ and where $L_y=30$.}
\label{dim}
\end{figure}

Fig.\ts\ref{dim} shows an example of the minimal dimension $N$ of $\mathcal{F}(A)$ with a fixed $L_y$ (big enough for the area law regime) when $\theta$ is far from the limits $\theta \rightarrow 0$ and $\theta\rightarrow\pi$, for the $\nu = 1$ state. As we can see, the minimal dimension of the matrix seems to be directly proportional to $\ell_x$.

For reference, in calculating the corner function for the von Neumann EE at a precision of $10^{-10}$, $\frac{L_y}{\ell_B} \gtrsim 30$ was more than enough for all angles for the $\nu = 1$ state. For the $\nu = 2$ state, the required length was also $\frac{L_y}{\ell_B} \gtrsim 30$ and for the $1^{st}$ filled LL, $\frac{L_y}{\ell_B} \gtrsim 35$. 
Also, generally, smaller angles don't require as a big a minimal length: for the $\nu = 1$ state, $\frac{L_y}{\ell_B} \gtrsim 15$ was enough for $\theta = 30^{\circ}$, whereas $\theta = 90^{\circ}$ and $\theta = 175^{\circ}$ required $\frac{L_y}{\ell_B} \gtrsim 25$. 

By and large, the results obtained for smaller angles have many more significant digits of precision, but require more computing power, as a bigger matrix is necessary to completely ``define'' the cut which is spatially larger than in the smooth limit. This is why we weren't able to compute the corner function for very small angles ($\theta \leq 0.035$). The results close to a smooth cut are much less precise. Indeed, the corner functions are small in that limit, quite close to machine precision. All computations in the large angle limit were however much less demanding. 

The precision of the constants for the asymptotic behavior near $\pi$ was limited by the fitting of the ratio of two very small functions (a($\theta \rightarrow \pi$) and $(\theta-\pi)^2$), whereas the precision of constants for the asymptotic behavior at small angle was limited by the fact that we weren't able to get data for the corner functions at very small angles.
 
\bibliographystyle{apsrev4-1}
\bibliography{bibliographie} 

\end{document}